\documentclass[sigconf]{acmart}

\usepackage{soul, booktabs, subcaption, makecell, multirow, tabularx, dsfont, pifont} 
\usepackage[short, nocomma]{optidef} 
\usepackage{enumitem}
\usepackage{natbib} 
\usepackage[colorinlistoftodos]{todonotes} 
\usepackage{cleveref} 
\usepackage{caption} 

\setcopyright{acmlicensed}
\copyrightyear{2025}
\acmYear{2025}
\acmDOI{XXXXXXX.XXXXXXX}
\acmConference[Conference EAAMO '25]{The fifth ACM Conference on Equity and Access in Algorithms, Mechanisms, and Optimization}{November 05--07,
  2025}{Pittsburgh, PA}

\begin{document}

\title{A community-driven optimization framework for redrawing school attendance boundaries}


\author{Hongzhao Guan}
\affiliation{%
  \institution{Georgia Institute of Technology}
  \city{Atlanta, GA}
  \country{USA}
  }
\email{hguan7@gatech.edu}

\author{Paul Riggins}
\affiliation{%
  \institution{Northeastern University}
  \city{Boston, MA}
  \country{USA}
  }
\email{p.riggins@northeastern.edu}

\author{Tyler Simko}
\affiliation{%
  \institution{University of Michigan}
  \city{Ann Arbor, MI}
  \country{USA}
  }
\email{tsimko@umich.edu}

\author{Jasmine Mangat}
\affiliation{%
  \institution{Northeastern University}
  \city{Boston, MA}
  \country{USA}
  }
\email{j.mangat@northeastern.edu}

\author{Cassandra Moe}
\affiliation{%
  \institution{Northeastern University}
  \city{Boston, MA}
  \country{USA}
  }
\email{moe.c@northeastern.edu}

\author{Urooj Haider}
\affiliation{%
  \institution{Northeastern University}
  \city{Boston}
  \country{USA}
  }
\email{lnu.ur@northeastern.edu}

\author{Frank Pantano}
\affiliation{%
  \institution{Winston-Salem/Forsyth County Schools}
  \city{Winston-Salem, NC}
  \country{USA}
  }
\email{fnpantano@wsfcs.k12.nc.us}

\author{Effie G. McMillian}
\affiliation{%
  \institution{Winston-Salem/Forsyth County Schools}
  \city{Winston-Salem, NC}
  \country{USA}
  }
\email{egmcmillian@wsfcs.k12.nc.us}

\author{Genevieve Siegel-Hawley}
\affiliation{%
  \institution{Virginia Commonwealth University}
  \city{Richmond, VA}
  \country{USA}
  }
\email{gsiegelhawle@vcu.edu}

\author{Pascal Van Hentenryck}
\affiliation{%
  \institution{Georgia Institute of Technology}
  \city{Atlanta, GA}
  \country{USA}
  }
\email{pvh@gatech.edu}

\author{Nabeel Gillani}
\affiliation{%
  \institution{Northeastern University}
  \city{Boston, MA}
  \country{USA}
  }
\email{n.gillani@northeastern.edu}

\renewcommand{\shortauthors}{Guan et al.}

\begin{abstract}
The vast majority of US public school districts use school attendance boundaries to determine which student addresses are assigned to which schools. Existing work shows how redrawing boundaries can be a powerful policy lever for increasing access and opportunity for historically disadvantaged groups, even while maintaining other priorities like minimizing driving distances and preserving existing social ties between students and families. This study introduces a multi-objective algorithmic school rezoning framework and applies it to a large-scale rezoning effort impacting over 50,000 students through an ongoing researcher-school district partnership. The framework is designed to incorporate feedback from community members and policymakers, both by deciding which goals are optimized and also by placing differential ``importance'' on goals through weights from community surveys. Empirical results reveal the framework's ability to surface school redistricting plans that simultaneously advance a number of objectives often thought to be in competition with one another, including socioeconomic integration, transportation efficiency, and stable feeder patterns (transitions) between elementary, middle, and high schools. The paper also highlights how local education policymakers navigate several practical challenges, like building political will to make change in a polarized policy climate. The framework is built using open-source tools and publicly released to support school districts in exploring and implementing new policies to improve educational access and opportunity in the coming years.
\end{abstract}





\maketitle

\section{Introduction}
In October 2023, Winston-Salem/Forsyth County Schools (WS/FCS) was awarded a ``Fostering Diverse Schools'' planning grant by the US Department of Education~\cite{jacobson2023fostering}\footnote{The school district partner is named with their permission to ground the project in a concrete historical and practical context.} to modernize its school attendance boundaries---the catchment areas that determine which student addresses are assigned to attend which schools---for the first time in over 30 years. WS/FCS is the fourth largest school district in North Carolina and the 81st largest in the US. However, the 50,000+ students who attend WS/FCS and its schools remain highly segregated by racial/ethnic background and socioeconomic status (SES). Across the United States, school segregation has long perpetuated achievement gaps~\cite{reardon2019geography} and prevented students and families from accessing social networks that might improve intergenerational income mobility and quality of life~\cite{chetty2022socialcapitalII}.

The two primary objectives of the district's effort to modernize attendance boundaries were (1) to improve transportation efficiency and (2) to balance SES integration. While race/ethnicity and SES are correlated---particularly when it comes to documenting academic achievement gaps~\cite{reardon2018testgaps}---they affect future life outcomes for young people differently~\cite{chetty2018race}. Still, the fraught legal landscape around designing student assignment policies to foster more racially/ethnically integrated schools~\cite{pics_kennedy}, coupled with evidence highlighting the benefits of disrupting concentrations of poverty in schools to achieve more cross-SES connections (a powerful notion of ``social capital''~\cite{chetty2022socialcapitalII}), made SES integration a project priority.

The district envisioned redrawing boundaries as a promising policy solution to advance both transportation efficiency and SES integration by incorporating both infrastructural (e.g., new roads) and demographic (e.g., diversifying student body) changes from the past three decades. While WS/FCS has an active within-district school choice system that enables students to attend schools they are not default-assigned to via boundaries (like magnet and other specialized programs), the vast majority (approximately 2/3) of students in the district attend their boundary-assigned schools.

Like WS/FCS, school districts across the United States use attendance boundaries to assign students to schools, and regularly redraw them to account for goals like population changes \citep{castro2024drawn}. While the lines between separate school districts contribute the most to demographic segregation, within-district lines---i.e., attendance boundaries---are powerful levers for fostering more integrated learning environments~\cite{fiel2013boundaries}. Accordingly, some large school districts, like WS/FCS's neighbor---Charlotte Mecklenburg Schools---have even established policies that require regular, comprehensive re-evaluation of (and possible updates to) attendance boundaries in order to keep pace with demographic, infrastructural, and other local changes~\cite{cms2025review}. As demographic and policy changes across the country increasingly impact school district enrollments, districts may face the need to change boundaries to improve operational efficiency---without sacrificing educational quality.

School districts like WS/FCS face common considerations when updating their attendance boundaries: typically, districts seek to incorporate (1) administrative constraints (e.g., school enrollment capacities), (2) existing policy design choices (e.g., ``feeder patterns" that connect students between elementary, middle, and high schools; ``compact'' boundaries for more efficient transportation routes), and (3) community feedback (e.g., feedback on proposed plans and differing priorities in map redesign) into their updated assignment plans. In practice, attendance boundary updates are manual processes carried out by demographers---but such processes can be cumbersome and prevent the identification of proposals that both adhere to administrative constraints and adequately balance across different goals. To account for these challenges, researchers have proposed many approaches, often based on integer programming, while accounting for administrative constraints in rezoning efforts~\cite{clarke1968, holloway1975interactive}. More recent work extends these foundational efforts to incorporate specific policy designs (such as school choice programs \citep{allman2022} or feeder patterns \citep{ozel2025community}), with calls to incorporate community feedback fairly, without privileging the ``loudest voices''\citep{gillani2023air}. Providing opportunities for meaningful community input is particularly important, as changes to existing attendance boundaries are important policy changes for students and families, and as a result can be politically contentious \citep{castro2024drawn}.

This paper presents the results of an ongoing researcher-practitioner partnership to simultaneously redraw 67 elementary, middle, and high school boundaries for more than 50,000 students. The algorithmic framework developed to support this effort is guided by the following four priorities---each of which constitutes a term in the redistricting algorithm's objective function: (1) transportation efficiency, (2) SES integration, (3) stable feeder patterns, and (4) compactness. The framework also incorporates administrative constraints, like school boundary capacities. The paper describes the iterative development of this framework through close collaboration with community stakeholders and local education officials, highlighting three particular contributions: 
\begin{enumerate}
    \item A novel optimization framework for the School Attendance Boundaries Redrawing (SABR) problem that jointly rezones multiple school levels (e.g., elementary, middle, and high schools) within a unified optimization model. Evaluating the framework through a field study reveals potential for redrawing boundaries in ways that could simultaneously improve driving distances, SES integration, and feeder pattern cohesion---thereby satisfying different (often competing) priorities. The proposed framework is a flexible, multi-objective approach with broad applicability beyond WS/FCS.
    \item The inclusion of feedback from community members and policymakers in two ways: 1) in the \textit{design} stage by allowing control over which goals are optimized, and 2) in the \textit{optimization} stage through the use of ``importance'' weights that control the degree to which different goals are optimized. These weights are estimated using over 8,000 survey responses from community members, adjusted to account for representational disparities in community engagement. 
    \item A real-world demonstration of a community-driven researcher-practitioner partnership at an unusually large scale. The collaboration is transparent and built upon open-source tools, enabling smooth adaptation by other researchers and districts. It also highlights how researchers and practitioners may work together to discover and implement modeling priorities, and ultimately, determine which algorithmically-generated policy drafts to release for public input.
\end{enumerate}
\noindent The remainder of the paper outlines how the study builds on existing work; formally introduces SABR and the optimization model; describes the collaboration with WS/FCS; and presents empirical results. It concludes with a series of takeaways on research partnerships, limitations, and suggestions for future work.
\section{Related Work}

The present paper builds on a long history of considering school redistricting as an optimization problem \citep[see, e.g., ][]{smilowitz2020}. Traditionally, school redistricting models have attempted to solve a student assignment problem given a set of constraints. For example, \citet{clarke1968} in 1968 aimed to find ``a plan of assignment of students to schools which achieves the desired ethnic composition at each school'' under a travel time constraint. Further, these approaches have long recognized various kinds of constraints involved in school redistricting: \citep{holloway1975interactive} categorize considerations as educational (e.g., teacher / student ratio), political (e.g., parental preferences, travel time, etc.), economic (e.g., fixed school capacities), and administrative (e.g., prohibit certain road crossings, policy-defined goals for student demographic composition). More recent efforts have also adapted this ``single-objective'' approach to contemporary policy questions, such as school choice \citep{allman2022}. Similarly, \citep{gillani2023redrawing} and \citep{simko2024school} both adopt the single-objective approach to examine how redrawing school attendance zones and school district boundaries, respectively, could reduce racial/ethnic segregation.

Other related work has adopted an explicitly multi-objective approach that attempts to optimize several criteria at the same time \citep{sutcliffe1984goal, diamond1987multiobjective, lemberg2000school, bouzarth2018assigning}. For example, \citet{bouzarth2018assigning} offer a model that seeks to optimize both student socioeconomic integration and transportation costs. In recent work, \citet{ozel2025community} introduce a stream-based approach, where students follow a predetermined sequence of schools intended to represent feeder patterns. For example, students in area X are assigned to ``Elementary School A'', ``Middle School B'', and ``High School C''. This stream-based approach builds feeder patterns into the preparation of the input data, which can improve computational efficiency by pre-constraining the feeder patterns allowed during optimization.

This paper builds on these existing efforts in both substantive and methodological ways. Substantively, the paper presents results of an ongoing partnership at redrawing school boundaries at scale: 67 boundaries simultaneously across more than 50,000 elementary, middle, and high school students. The present approach is designed to incorporate community feedback into the model to select both optimization priorities and the relative importance of those priorities through model weights. Methodologically, this study extends existing work by adopting a multi-objective approach that explicitly optimizes the feeder patterns that connect each school level. While conceptually related to the ``streams'' approach presented in \citep{ozel2025community}, it optimizes school levels in a multi-objective way, instead of embedding certain constraints/objectives into data preparation. This reduces the need for researchers/practitioners to pre-identify possible feeder patterns---which, while possibly a valuable exercise---may be time consuming or otherwise impractical if the space of possible feeder patterns is large (as it is in this paper's case study). Finally, this paper's use of the open-source CP-SAT solver \cite{cpsat} allows school districts and partners to directly use the presented approach without relying on commercial tools.
\section{The School Attendance Boundaries Redrawing (SABR) Problem}
\label{sect:problem}
This section presents the School Attendance Boundaries Redrawing (SABR) problem, which serves as the backbone of this paper. The main goal of SABR is to develop a zoning plan for a public school district that achieves multiple goals (e.g., reducing transportation distances; SES integration; etc.) across several school levels (e.g., elementary, middle, and high schools) simultaneously---all while satisfying different constraints. Although this paper is grounded in the collaboration with WS/FCS, SABR is general and has broad applicability across numerous districts in the United States, and potentially on a global scale. Table~\ref{tab:nomenclature} summarizes the paper's notation.

\subsection{Community-Informed Optimization}
There is a long history of using optimization methods in public education planning~\cite{smilowitz2020}, many of which factor in community considerations. This paper defines ``community'' as 1) those affiliated with schools who would be most directly affected by policy changes (e.g., students, families, and teachers), and 2) those responsible for designing and enacting new policies (e.g., district staff). The framework incorporates these groups' perspectives to inform two key decisions: a) \textit{what} the model's guiding priorities and constraints should be, and b) \textit{how} those priorities and constraints should be weighed against one another in a multi-objective setting. Additionally, given the perennial challenge of obtaining representative feedback from different demographic groups in community engagement settings~\cite{le-dantec2015strangers}, the framework also seeks to account for disparities in \textit{whose} input gets to shape redistricting goals and inputs. The next sections present objectives and constraints informed by community engagement through WS/FCS, but also prevalent in other redistricting contexts~\cite{overney2025boundarease}. These modeling decisions are presented in a general fashion through technical notation in Section~\ref{sect:cp_model_big}, and then Section~\ref{sect:case_study} offers more details on how they were derived from community input while seeking to account for representational disparities as much as possible.

\subsection{Problem Definition}
\label{subsect:problem}
SABR aims to assign a set of geographically-defined ``planning units'' in a district, denoted as set $\mathcal{P}$, to schools in a way that minimizes specific metrics, while satisfying several constraints. Planning units can be widely-recognized entities like Census Blocks or Block Groups, or customized units tailored to a particular task, such as those used in the case study of this paper (see Section~\ref{subsect:data}). 

The set of schools, $\mathcal{SH}$, can be partitioned into multiple mutually-exclusive subsets based on school level $l \in \mathcal{L}$. By default, $\mathcal{L} = \{ 1, 2, 3 \}$,  represents elementary, middle, and high schools, respectively, reflecting the most common configurations seen in the United States. However, the definition of $\mathcal{L}$ can be adapted to different grade structures based on the characteristics of the school district. For example, K-4 may be treated as one level, while grades 5-8 may constitute another.

Using $l$, the set $\mathcal{P}$ can also be partitioned into multiple mutually exclusive subsets, as different school levels may utilize different groups of planning units. In some cases, planning units and schools may appear multiple times at different levels: for instance, certain schools serve both middle and high school students, and some planning units apply to all school levels. Despite this overlap, SABR assigns distinct integer identifiers to partition $\mathcal{P}$ and $\mathcal{SH}$ into mutually exclusive sets at different level, ensuring clarity even when the entities remain the same. That is to say, $p \in \mathcal{P}^l$ and $s \in \mathcal{SH}^l$ inherently contain level information, making it unnecessary to use $p^l$ or $s^l$ for further clarification.

Students at each level, represented by $n \in \mathcal{N}^l$, have residential addresses that determine their corresponding planning units ($\bar{p}_n^l$). Unlike $p$ and $s$, $l$ is necessary here because a student's residential planning unit may vary across different school levels. Once the ``planning unit $\rightarrow$ school'' assignments are determined, students are expected to attend their designated schools based on $\bar{p}_n^l$ (a simplifying assumption relevant to this paper's case study; this can be augmented to account for school choice, e.g.~\cite{guan2025rwc}). Bar notation throughout the paper indicates constants derived from data. 

With the preliminary concepts established, Section~\ref{sect:cp_model_big} presents the technical details of modeling SABR. Readers primarily interested in practical applications may proceed directly to Section~\ref{sect:case_study}, which showcases how a U.S. school district collaborates with researchers to implement SABR through a field deployment.
\section{Constraint Programming for SABR}
\label{sect:cp_model_big}

The optimization model presented in this paper uses Constraint Programming (CP) ~\cite{van1992constraint}, which offers strong capabilities for handling complex logical expressions. Logical constraints can be formulated directly without the need for additional variables, as is often necessary in Mixed-Integer Programming. This flexibility makes Constraint Programming particularly well-suited for modeling SABR.

Figure~\ref{fig:cp} presents the constraint programming model. Users can choose the objectives they want to optimize by setting $\mathcal{O}$. Then the model should include, as constraints, all appropriate equations and logical expressions related to objectives in $\mathcal{O}$. The rest of this section details the decision variables, objective functions, calibrations, and constraints involved.

It is important to highlight that the CP model presented here is highly general, consistent with the SABR problem definition in Section~\ref{sect:problem}. Although developed through collaboration with WS/FCS, the model is broadly applicable to many school districts. This section also demonstrates this point by explaining how this general CP model can be adapted to accommodate district-specific requirements. In Section~\ref{sect:case_study}, this paper provides a real-world case study showing how the research team modified, applied, and deployed a tool based on this CP model for a specific district.

\subsection{Decision Variables and ``Zonings''}
After assigning integer identifiers to each $s \in \mathcal{SH}$, the integer variable $y_p$ represents the school to which planning unit $p$ is assigned. This model additionally uses a binary variable representation $z_{p, s}$ to indicate whether planning unit $p$ is assigned to school $s$. The following logical expression links $y$ and $z$ and ensures each $p$ is zoned to exactly 1 school:
\begin{equation}
\label{eq:link_y_z}
    y_p = s \leftrightarrow z_{p, s} \qquad \forall s \in \mathcal{SH}, p \in \mathcal{P}
\end{equation}
Note that $z_{p, s}$ is trivially 0 whenever $p$ and $s$ correspond to different levels $l \in \mathcal{L}$. Those variables can be safely omitted from the model.

The study incorporates these two representations because of their varying degrees of effectiveness when building Constraint Programming models, which will become clear later. These decision variables are collectively organized into the vectors $\mathbf{y}$ and $\mathbf{z}$, which is referred to as a ``zoning'' in this paper. The status-quo zoning refers to the existing zoning established by the district before running the optimization model. The status-quo zoned school for planning unit $p$ and student $n$ are $\bar{s}_p$ and $\bar{s}_n$, respectively.

\begin{table}[t]
\footnotesize
\centering
\begin{tabularx}{\columnwidth}{l X}
\toprule
\textbf{Notation} & \textbf{Definition} \\
\midrule
\textbf{Sets}: & \\
$\mathcal{L}$ & School level or simply ``level''. By default, $\mathcal{L} = \{ 1, 2, 3 \}$, representing elementary, middle, and high, respectively. $\mathcal{L}$ can also be customized to the needs of school districts. \\
$\mathcal{P}$, $\mathcal{P}^l$ & All school district planning units and planning units for school level $l$. Each planning unit $p \in \mathcal{P}^l$ is assigned a unique integer ID. The sets $\mathcal{P}^l$ are disjoint. \\
$\mathcal{SH}$, $\mathcal{SH}^l$ & All schools and schools at level $l$. Each school $s \in \mathcal{SH}^l$ is assigned a unique integer ID. The sets $\mathcal{SH}^l$ are disjoint. \\
$\mathcal{N}, \mathcal{N}^l, \mathcal{N}_p$ & All students, all students currently at school level $l$, and all students who currently reside in planning unit $p$. \\
$\mathcal{G}, \mathcal{G}^l, \mathcal{G}_p$ & All students in a specific group, those currently at school level $l$, and those who currently reside in planning unit $p$. \\
\textbf{Decision Vars.}:\hspace{-3pt} & \\
$y_p$ & Integer variables represent the school to which planning unit $p$ zoned. \\
$z_{p, s}$ & Binary variables indicate if planning unit $p$ is zoned to school $s$.\\
$o_{s}$, $o^\text{g}_s$ & Integer variables indicate the number of students and the number of students in a specific group zoned to school $s$, dynamically based on $z_{p, s}$. \\
$d_{s}$ & Dissimilarities of school $s$, based on $o_{s}$ and $o^\text{g}_s$. \\
$u_{p_1, p_2}$ & Binary variables indicate if an edge-cut for planning unit $p_1$ and $p_2$ is performed. \\
$f_{s_1, s_2}$ & Binary variables indicate the existence of a feeder-pattern from a lower level school $s_1$ to a higher level school $s_2$. \\
\textbf{Parameters}: & \\
$\mu_n^\text{ratio}$ \enspace $\mu_n^\text{abs}$ & The maximum ratio or absolute difference on travel distance increment for student $n$.\\
$o_s^{\max}$ \enspace $o_s^{\min}$ & The upper and lower limits on the number of students that may be zoned to each school. \\
$o_s^{*}$ & The desired number of students that should be zoned to school $s$. \\
$\epsilon$ & Threshold on feeder-pattern \\
$\lambda$ & Margin-value for dissimilarity \\
$w_{s, obj}$ & Weights assigned to each school and each objective in the optimization model. \\
$b_{l, obj}$ & Calibrations assigned to each level and each objective in the optimization model. \\
\textbf{Constants}: & \\
$\bar{l}_{\max}$ & The maximum school level. \\
$\bar{a}_{n, s}$ & Estimated travel distance from student $n$'s residential address to school $s$. \\
$\bar{p}_n^l$ & The planning unit that student $n$ resides in at school level $l$. Note that the planning unit in which a student resides at different school levels may differ.\\
$\bar{s}_p$ & The status-quo zoned school for planning unit $p$. \\
$\bar{s}_n$ & The status-quo zoned school for student $n$.\\
$\bar{o}_s, \bar{o}^\text{g}_s$ & The number of students and the number of students in a specific group zoned to school $s$ in the status-quo zoning. \\
\bottomrule
\end{tabularx}
\caption{Notation used throughout this study.}
\label{tab:nomenclature}
\end{table}
\captionsetup[figure]{font=Large,labelfont=bf}
\begin{figure}[!t]
\resizebox{\columnwidth}{!}{%
\large
\begin{minipage}{0.6\textwidth}
\begin{mini!}
    {}
    {   
        \text{Objective}(\mathcal{L})
    }
    {\label{formulation:cp_c}}
    {}
    \addConstraint
    {
       \text{Select From} \quad
       \eqref{eq:link_y_z}, \eqref{eq:zoned_population}, 
       \eqref{eq:zoned_population_group},
    }
    {   
       \eqref{eq:difference_to_district},
       \eqref{eq:edge_cut}, 
       \eqref{eq:a_feeder_pattern}
       \notag
    }
    {
        \quad \text{Based On } \mathcal{O} 
    }
    \addConstraint
    {
       \sum\limits_{s \in \mathcal{SH}^l} z_{\bar{p}_n^l, s} \cdot \bar{a}_{n, s} 
    }
    {   
        \leq (1 + \mu_n^\text{ratio}) \cdot \bar{a}_{n, \bar{s}_n}
        \label{formulation:constraint_travel_distance_ratio}
    }
    {
        \quad \forall l \in \mathcal{L}, p \in \mathcal{P}^l, n \in \mathcal{N}_p
    }
    \addConstraint
    {
        o_s^{\min} \leq o_s
    }
    {   
        \leq o_s^{\max}
        \label{formulation:constraint_school_capacity}
    }
    {
        \quad \forall l \in \mathcal{L}, s \in \mathcal{SH}^l
    }
    \addConstraint
    {
        \text{Ensure Contiguity}(p)
    }
    {   
        \label{formulation:constraint_contiguity}
    }
    {
        \quad \forall l \in \mathcal{L}, p \in \mathcal{P}^l
    }
    \addConstraint
    {
        y_p
    }
    {   
        \in \mathcal{SH}^l
        \label{formulation:y}
    }
    {
        \quad \forall l \in \mathcal{L}, p \in \mathcal{P}^l
    }
    \addConstraint
    {
        z_{p, s}
    }
    {   
        \in \{0, 1\}
        \label{formulation:z}
    }
    {
        \quad \forall l \in \mathcal{L}, p \in \mathcal{P}^l, s \in \mathcal{SH}^l
    }
\end{mini!}
\vspace{-5pt}
\caption{General Constraint Programming Model for SABR}
\label{fig:cp}
\Description{The Constraint Programming Model.}
\end{minipage}
}
\end{figure}
\captionsetup[figure]{font=normalsize,labelfont=bf}
\subsection{Objective Function}
\label{subsect:objective_function}
\begin{equation}
\label{eq:objective}
\resizebox{0.9\columnwidth}{!}{
$
\begin{aligned}
       Objective(\mathcal{L}) = \quad
       & \sum\limits_{l \in \mathcal{L}} 
       \sum\limits_{{obj} \in \mathcal{O}} 
       b_{l, obj} \cdot \mathbf{f}_{obj}(l) \\
       & \text{where } \mathcal{O}  \text{ represents the selected objectives} \\
\end{aligned}
$
}
\end{equation}
SABR is modeled by multi-objective optimization that uses a weighted sum approach. The objectives outlined in this paper stem from geographic, social science, and transportation considerations, which encapsulate a wide range of factors. Each objective is deliberately designed based on discussions with stakeholders and insights from previous literature. However, researchers and policymakers retain the flexibility to only select some of the objectives  or incorporate additional terms as needed. To aggregate these terms, it is important to account for their differing scales. Therefore, calibrations $b_{l, obj}$ are introduced to ensure the model is able to adequately balance and trade-off among them (see \Cref{sec:calibration}). Equation~\eqref{eq:objective} presents the objective function, and the remainder of this section explains each objective in detail. Additionally, weight values (e.g., $w_{s, distance}$) used in each objective for each school are also derived from empirical surveys, and will be discussed later (see \Cref{sec:weights}).

\subsubsection{Travel Distance Ratio}
Parents and students generally prefer to be assigned to schools close to their homes. This objective compares the sum of travel distances for all students assigned to school $s$ to the total travel distance of the same group of students to their originally-assigned schools. Here, $\bar{a}_{n, s}$ denotes the travel distance from a student's household to a particular school.

\begin{equation}
\label{eq:objective_distance}
    \mathbf{f}_{distance}(l) = \sum\limits_{s \in \mathcal{SH}^l}
    w_{s, distance}
    \cdot
    \cfrac{
        \sum\limits_{n \in N^l} \bar{a}_{n, s} 
        \cdot 
        z_{\bar{p}_n^l, s}
    }{
        \sum\limits_{n \in N^l} \bar{a}_{n, \bar{s}_n} 
        \cdot
        z_{\bar{p}_n^l, s}
    }
\end{equation}

\subsubsection{Boundary Capacity}
Ensuring school boundaries do not contain too many or too few students is essential for fostering effective learning environments. 

Integer variable $o_s$ represents the number of students zoned to each school, where $\mathcal{N}_p$ includes all students who currently reside in planing unit $p$.
\begin{equation}
\label{eq:zoned_population}
    o_s = \sum\limits_{p \in \mathcal{P}} z_{p, s} \cdot |\mathcal{N}_p| \qquad \forall s \in \mathcal{SH}
\end{equation}
The objective function for boundary capacity is then defined as:
\begin{equation}
    \mathbf{f}_{population}(l)
    = 
    \sum\limits_{s \in \mathcal{SH}^l} 
    w_{s, population}
    \cdot
    \Big| 1 - \cfrac{o_s}{o_s^*} \Big|
\end{equation}
where  $o_s^*$ denotes the desired capacity for school $s$.
\subsubsection{Demographic Balancing} 
Ensuring demographic balance (SES integration in the context of the district collaboration) in schools can help provide equitable access to educational resources and diverse social interactions, which are linked to better academic and social outcomes for all students~\cite{johnson2011desegregation,chetty2022socialcapitalII}.

For a target group $\mathcal{G} \subseteq \mathcal{N}$ (e.g., students from lower socioeconomic status (SES) backgrounds), the subset of these students within each planning unit $p \in \mathcal{P}$ is represented as $\mathcal{G}_p$. Similar to $o_s$, the variable $o^\text{g}_s$, represents the number of students in $\mathcal{G}$ zoned to each school $s$ and is defined as:
\begin{equation}
\label{eq:zoned_population_group}
o^\text{g}_s = \sum\limits_{p \in \mathcal{P}} z_{p, s} \cdot |\mathcal{G}_p| \qquad \forall s \in \mathcal{SH}
\end{equation}
For a school $s$ in level $l$, the following equation measures how evenly students in $\mathcal{G}^l \subseteq \mathcal{N}^l$ is distributed across schools compared to their overall proportion in the district. It calculates the difference between a school's share of the target group and the group's overall representation at the same school level. $d_s$ is closely related to the popular ``dissimilarity'' measure of segregation~\cite{massey1988segregation}; it is modified here to produce a per-school measure (dissimilarity measures segregation at a system-wide level, e.g., the whole school district).
\begin{equation}
\label{eq:difference_to_district}
    d_s = \bigg| \enspace
        \cfrac{
            o^\text{g}_s
        }{
            o_s
        }
        -
        \cfrac{
            |\mathcal{G}^l|
        }{
            |\mathcal{N}^l|
        }
    \enspace \bigg|
    \qquad \forall l \in \mathcal{L}, s \in \mathcal{SH}^l
\end{equation}  
Lastly, the objective function for demographic balancing is defined as the following. Note that this objective function does not directly work with $d_s$, instead, a margin value $\lambda$ is introduced. If a school's demographic proportion is within $\lambda$ of the district-level proportion, the difference is effectively ignored. This prioritizes schools that are significantly imbalanced, and draws inspiration from prior court cases about desegregation like~\cite{swann1971charlotte}, after which lower court(s) required that school enrollments across demographic groups must fall within some margin of district-wide demographics.
\begin{equation}
    \label{eq:objective_balance}
    \mathbf{f}_{balance}(l) = 
    \sum\limits_{s \in \mathcal{SH}^l}
    w_{s, balance}
    \cdot
    \max(d_s, \lambda)
\end{equation}

For reference, the aforementioned system-wide dissimilarity measure could be similarly defined as:
\begin{equation}
    \mathbf{f}_{balance-d}(l) = \cfrac{1}{2}
    \sum\limits_{s \in \mathcal{SH}^l}
    w_{s, balance-d}
    \cdot
    \bigg|
    \cfrac{o^\text{g}_s}{|\mathcal{G}^l|}
    -
    \cfrac{o_s - o^\text{g}_s}{|\mathcal{N}^l| - |\mathcal{G}^l|}
    \bigg|
\end{equation}

\subsubsection{Compactness} Compactness in redistricting plans can help ensure geographic cohesiveness, making it easier to understand and manage for parents and policymakers. Compact regions can also prevent irregularly shaped boundaries that may lead to inefficiencies or unintended unfairness---which partially explains their importance as a factor when designing legislative districts~\cite{gurnee2021fairmandering}. 

This paper adopts the concept of ``edge-cut compactness'' from \cite{dube2016beyond}. In this context, planning units in $\mathcal{P}^l$ collectively form an undirected graph, where each planning unit represents a vertex, and an edge connects two vertices if the corresponding planning units are adjacent. The zoning process partitions this graph, so that the planning units assigned to each school construct disjoint subgraphs. Minimizing edge-cuts promotes compact regions.

Based on the graph formed from $\mathcal{P}^l$, a new set $\mathcal{A}^l$ is defined to include all adjacent pairs of planning units. Next, a binary variable $u_{p_1, p_2}$ is introduced to indicate whether an ``edge-cut'' is needed. That is to say, an ``edge-cut'' is performed when two adjacent planning units are assigned to different schools. 
\begin{equation}
\label{eq:edge_cut}
    u_{p_1, p_2} \leftrightarrow (y_{p_1} != y_{p_2}) \quad \forall (p_1, p_2) \in \mathcal{A}^l
\end{equation}
The compactness function for level $l$ is then defined as:
\begin{equation}
    \label{eq:objective_compactness}
    \mathbf{f}_{compact}(l) = \sum\limits_{(p_1, p_2) \in \mathcal{A}^l} u_{p_1, p_2}
\end{equation}
The primary advantage of this method for measuring compactness is that it is computationally efficient within the optimization framework, as it does not require overly complex formulations (e.g., the calculation of zone perimeters or bounding boxes).

\subsubsection{Feeder Patterns} A student's feeder pattern represents the sequence of elementary, middle, and high schools they are assigned to attend (in essence, how the attendance boundaries of different schools at each of these levels ``overlap''). Feeder patterns are important factors in SABR because they can help keep peer groups together as they transition between school levels, which can enhance social stability and impact the learning experience.

This objective aims to quantify the number of feeder patterns, which represent distinct transitions students would experience vis-a-vis their boundary-assigned schools when transitioning from lower to higher school levels. A new binary variable $f_{s_1, s_2}$ is introduced to represent the existence of a specific feeder pattern. It is set to 1 if there exist enough students who are assigned to $s_1$ at level $l_1$ and, based on their residential address, would be assigned to $s_2$ at level $l_2$ in the future, where $l_1 < l_2$. Otherwise, $f_{s_1, s_2}$ is set to 0.
\begin{equation}
\label{eq:a_feeder_pattern}
    f_{s_1, s_2} = \mathds{1} \bigg[
        \sum\limits_{n \in \mathcal{N}^{l_1}}
         z_{\bar{p}_n^{l_1}, s_1}
         \cdot
         z_{\bar{p}_n^{l_2}, s_2} 
         \geq \epsilon
    \bigg]
\end{equation}
It is important to highlight that $f$ must be determined at the student level rather than the planning unit level. This is because the planning unit in which a student resides at school levels $l_1$ and $l_2$ may differ; they could be the same or partially overlap. The parameter $\epsilon$ is set to 1 by default but can be assigned a higher threshold value to ensure that relatively infrequent transition sequences are not counted as distinct feeder patterns.

A new term can now be introduced in the objective function, which counts the number of feeder patterns that exist between school levels $l$ and $l + 1$:
\begin{equation}
\label{eq:objective_feeder}
\resizebox{0.8\columnwidth}{!}{
$
    \mathbf{f}_{feeder}(l) =  
    \begin{cases}
    0  & \text{if } l = \bar{l}_{\max} \\
    \sum\limits_{s_1 \in \mathcal{SH}^{l}} w_{feeder, s_1} \cdot \sum\limits_{s_2 \in \mathcal{SH}^{l + 1}} f_{s_1, s_2} & \text{o.w}
    \end{cases}
$
}
\end{equation}

This objective is particularly interesting because it involves interactions across multiple school levels. Decisions made at one level can impact zoning at other levels. This interdependence represents one of the key innovations of the optimization model, and the reason that boundaries at all levels must be redrawn simultaneously (which increases the scale and complexity of the problem at hand).

\subsection{Calibration}
\label{sec:calibration}

Each of the terms in the objective function captures a qualitatively different concern as a nondimensionalized numerical value. The typical variations of the components $\mathbf{f}_{obj}(l)$ above might have radically different scales, depending on their construction and the underlying data. Performing an optimization requires balancing these in a sensible way. For instance, how should the objective judge between improvements in travel distance vs demographic integration? If integration can improve by 10\% but that would require an increase in the net travel distance of 100 miles, is that tradeoff ``worth it''? If it is, then the objective should decrease under this scenario; otherwise, it should increase.

To balance between objectives, this paper introduces calibration scalings $b_{l, obj}$ for each objective at each level. The goal is to choose scalings such that all $b_{l, obj} \cdot \Delta \mathbf{f}_{obj}(l)$ are comparable, where $\Delta \mathbf{f}$ is the change in an objective from status-quo after the zoning process (see Eq.~\eqref{eq:objective}).

Each objective term itself has several terms: compactness sums over an adjacency graph, while other objectives sum over individual schools. This yields $\Delta \mathbf{f}_{obj}(l) = N_{l, obj}\, \delta_{l, obj}$, where $N_{l, obj}$ is the number of terms and $\delta$ is the average change per term. A natural choice would be $b_{l, obj} = (N_{l, obj}\, \delta_{l, obj})^{-1}$. However, competing changes within an objective-level can sometimes cause fine-tuned cancellations of $\delta_{l, obj}$, even though the per-term (e.g., per-school) variation is larger. These cancellations are unlikely to persist after the introduction of per-school community weights (see \Cref{sec:weights}), which were not included in the calibration process to avoid risk of ``calibrating out'' real community preferences between objectives.

The study therefore sets $b_{l, obj} = (N_{l, obj}\, |\delta|_{l, obj})^{-1}$, where $|\delta|_{l, obj}$ is the average of the absolute values of the terms for objective $obj$ at level $l$. Across various scenarios, $|\delta|_{l, obj}$ is typically larger than $\delta_{l,obj}$ by a factor of a few, but sometimes up to an order of magnitude. The difference varies by objective and level.

Note that computing calibration scalings per level also enables balancing different numbers of terms at each level: there are many more elementary schools than high schools, for instance, but the algorithm should give each of those levels comparable weight in the overall objective ~\eqref{eq:objective}.

Finally, a specific zoning scenario must be chosen to calculate the values $|\delta|_{l,obj}$. One option would be to iteratively run the full multi-objective simulation and iteratively update $b_{l,obj}$, hoping to see the values converge. This would correspond to calibrating for comparable \textit{achieved progress} over all objectives. Though this may have merit, it is computationally expensive, and would warrant an analysis of its convergence properties (beyond the scope of this study). Instead, the team chose to calibrate based on \textit{achievable progress} across each objective individually: it computes $|\delta|_{l,obj}$ based on the results of single-objective runs for each objective (see \cref{subsect:experiment}). In each single-objective run, $b_{l,obj}$ is set to $1$ for the target objective and $b_{l,obj} = 0$ for all other objectives.

All single-objective runs were subject to the same travel distance, capacity, and contiguity constraints (see Section~\ref{subsect:cp_constraints}). The demographic balancing single-objective run was subject to an additional constraint, specifically relevant to that objective. See \Cref{subsect:experiment} for further details.

\subsection{Constraints}
\label{subsect:cp_constraints}
Users can choose the objectives they want to optimize by setting $\mathcal{O}$, and then select relevant equations to include in the model as constraints. For instance, constraint \eqref{eq:link_y_z} must be included along with the Compactness objective, but is otherwise not needed (since no other objective uses the vector $\mathbf{y}$). Regardless of the makeup of $\mathcal{O}$, the model also enforces these three key constraints:
\begin{itemize}[label=---]
    \item Travel Distance Increases: Constraint~\eqref{formulation:constraint_travel_distance_ratio} ensures that, for each student, driving distance to a newly-zoned school does not exceed some relative threshold in relation to the student's distance to their status quo-assigned school. The focus in this study is on driving distance (not time, as in prior work~\cite{gillani2023redrawing}) to minimize variance introduced by other factors like traffic, but the model can easily incorporate time instead. Moreover, although not explicitly included in the model, it is possible to define an absolute threshold for each student, $\mu_n^\text{abs}$, to ensure that the increase in travel distance remains within a reasonable range, especially when the original travel distance is already large and therefore even a moderate relative increase may be unpalatable.
    \item School Capacity Ranges: Constraint~\eqref{formulation:constraint_school_capacity} dynamically bounds the number of students zoned to each school. This study treats these bounds as input parameters.
    \item Contiguity: Constraint~\eqref{formulation:constraint_contiguity} maintains reasonable zoning shapes by ensuring that planning units assigned to a school that were contiguous in status-quo remain contiguous in any new zoning. This paper adopts the definition of contiguity constraints introduced in \cite{mehrotra1998contiguity}.
\end{itemize}
The current model is presented in a general form. Constraints~~\eqref{formulation:constraint_travel_distance_ratio}, \eqref{formulation:constraint_school_capacity}, and \eqref{formulation:constraint_contiguity} are enforced since travel distance, capacity, and contiguity are fundamental and intuitive to the SABR problem. However, additional constraints can be included as needed. For example, in this paper's case study, the dissimilarities $d_s$ are bounded by \eqref{formulation:constraint_dissimilairty_bound} for each school when Demographic Balancing is involved in the $Objective(\mathcal{L})$. Similarly, constraints can be added to limit feeder patterns. (See case study discussion in \Cref{subsect:experiment,subsect:stakeholder}.)

It is worth mentioning that Constraint~\eqref{formulation:constraint_travel_distance_ratio} ensures that a planning unit cannot be assigned to a school beyond a certain travel distance. To reduce the size of the problem and improve computational efficiency, the team removed this constraint from the model and instead handled it as a preprocessing step by eliminating all decision variables $z_{p,s}$ where $p$ is too far from school $s$ under specific $
\mu_{n}^\text{ratio}$ values. 
\section{Case Study}
\label{sect:case_study}
\subsection{Background}
The project team collaborated with Winston-Salem/Forsyth County Schools (WS/FCS) in North Carolina, USA to develop and apply the model presented in the previous section. WS/FCS is the fourth largest school district in North Carolina and 81st largest in the US, serving over 50,000 students. Racially and ethnically, it is diverse, with approximately 1/3 of the students classified as White, Black, and Hispanic/Latinx, respectively. The district is also politically diverse, as reflected by split voting patterns in the 2024 US Presidential Election\footnote{\url{https://er.ncsbe.gov/?election_dt=11/05/2024&county_id=34&office=ALL&contest=0}.}, and a 5-4 Democratic majority on the current School Board\footnote{\url{https://ballotpedia.org/Winston-Salem/Forsyth_County_Schools,_North_Carolina,_elections}.}. WS/FCS has a long history of desegregation, busing students post Brown vs. Board of Education, only to stop upon the formal end of court-ordered desegregation mandates and introduction of school choice policies in the 1990's~\cite{wsfoundation}. This transition also marked the last time the district meaningfully updated its school attendance boundaries. Therefore, when WS/FCS was notified that it had received a \$1M, 2-year planning grant through the Federal ``Fostering Diverse Schools'' program~\cite{jacobson2023fostering} to modernize its residential boundaries, the community's response included both hope for positive change, and fear due to uncertainty about how changes might affect one's own children; home values; and other factors. 

Importantly, WS/FCS also has an active within-district school choice program. Students can attend any school in the district, and are provided transportation to schools that fall within the same ``Zone of Choice'' (ZOC - a cluster of several schools, typically with boundaries that are adjacent to one another) as their default boundary-assigned school. Despite this program, approximately 2/3 of students in the district do not exercise school choice and instead attend their boundary-assigned schools. Unlike other larger school districts, WS/FCS does not use a centralized matching algorithm like deferred acceptance, top trading cycles, or other mechanisms to match students to over-subscribed schools. Instead, students are assigned priorities based on their residence (e.g., for ZOC schools); where their siblings have attended schools; and whether they have a parent who also works at a particular school as a staff member. At oversubscribed schools, these students are offered seats first (typically their are enough seats to accommodate them), and the remainder of applicants are offered seats based on a random lottery. 

Given that the vast majority of WS/FCS students attend their boundary-assigned schools, and that the district's focus for the FDS grant was on increasing SES integration and improving transportation vis-a-vis attendance boundaries, all modeling and analyses were conducted based on a child's boundary-assigned---not attended---school. This means that estimates of SES integration of a particular school, for example, are defined in terms of the students zoned to that school and not necessarily those who attend. Prior research has explored forecasting school choice~\cite{pathak2017demand}---including in the context of redrawing attendance boundaries~\cite{guan2025rwc}---both of which in theory could be applied in this setting to estimate school attendance post-rezoning. However, given the complexities of anticipating school choice, the district opted to first focus on ``resetting defaults''---i.e., attendance boundaries---before estimating and factoring in the extent to which families may or may not adhere to these new defaults, which will also constitute an important part of future planning processes before implementing new boundary policies. Practically, this meant that only those schools with attendance boundaries (67 of the 81 total schools) were included in this study.

At the time of writing this paper, a first draft of algorithmically-generated maps had been released for initial community input, and more than 4,000 feedback responses had been received in the first several weeks. The rest of this section describes the process for producing and releasing these maps, along with several experimental runs to evaluate the model and understand its behavior. Not all of these runs were released for community input and are instead included for empirical evaluation purposes. Future research will investigate the community's responses to maps and the ensuing process for iterating upon and releasing updated drafts.

\subsection{Datasets}
\label{subsect:data}
\subsubsection{Student enrollment}
Through both University IRB and data usage agreements, the school district provided de-identified student zoning and enrollment data for the 2024-2025 school year. This data included both the default-assigned and attended schools for each student in the district, along with their places of residence. The district also provided the locations and current attendance boundaries for each of the 67 schools.

\subsubsection{Socioeconomic Status (SES)}
Despite its prominent use across the social sciences, socioeconomic status (SES) is an ill-defined measure with varying definitions. It is often interpreted synonymously with income; yet income does not typically encapsulate the different types of disadvantage communities might face. Typical measures like Free and Reduced-Price lunch were not available at the student level to the project team due to privacy reasons; and even if they were, using them as a primary indicator of SES would suffer from a number of limitations~\cite{harwell2010frl}.

Therefore, the team started by taking inspiration from other school districts like Chicago Public Schools (CPS)~\cite{quick2016cps} to produce a geographically-defined measure of SES. A primary limitation of a geographically-defined measure is that students are classified based on where they live, even if their individual/family-level demographics differ from their surrounding areas. A benefit of such a measure, however, is that it enables the incorporation of different household-level variables that are available through the Census and American Community Survey.

Based on CPS' approach, the team started with five variables available through the American Community Survey at the Block Group Level: Median Household Income, Home Ownership Rates, Adult Educational Attainment (\% with a Bachelor's Degree), English Proficiency (\% only speaking English), and Family Composition (\% of dual-parent households). Similar to prior work~\cite{guan2025rwc}, the team then standardized each variable (z-scored) across all block groups in the district, and then averaged the z-scores across these different variables and z-scored this average to produce a standardized index measure. Block groups in the bottom third of the SES distribution were classified as lower-SES, and those in the middle and top thirds were classified as medium and higher-SES, respectively. Under this definition, approximately 40\% of students across the district were classified as lower-SES (suggesting that more densely-populated Block Groups were more likely to be classified as lower-SES). Note: the most salient difference in SES designations for this study is between lower and non-lower-SES areas, as the demographic balancing term in the redistricting model aims to reduce segregation according to this dichotomous notion of SES.

In reviewing the results with the district, however, an interesting finding emerged: there were some areas---particularly those in the rural parts of the district that were more likely to serve lower-income White students---that were not classified as lower-SES, despite serving a large fraction of lower-income students according to their Title I designation. This led the team to develop two additional SES measures based on the above method: one that included only Median Household Income (a typical coarse measure of SES), and another that computed an index including only Median Household Income and Adult Educational Attainment. Students were then classified as lower-SES if they lived in an area that fell in the bottom third of the z-score distribution for \textit{any} of these three metrics. Applying this more conservative definition yielded approximately 50\% of students in the district as being classified as lower-SES, and upon comparison with other district benchmarks like Title I status per school, aligned more closely to district leaders' expectations.  

\subsubsection{Planning units and contiguity}
An important decision in the redistricting data preparation process involved defining the geographic ``pieces''---i.e. planning units---that would form the basis of the model's re-assignment problem. Prior work has used Census Blocks~\cite{gillani2023redrawing,simko2024school,guan2025rwc}, which are the smallest administrative units released by the US Census. Other studies have also used larger units, like Census Block Groups~\cite{allman2022}. These are approximately 6,300 Census Blocks that comprise WS/FCS, and just under 300 Census Block Groups---suggesting large differences in the geographic resolution of these two units. Perhaps most crucially, neither unit encapsulates notions of ``neighborhood'' or ``community''---i.e., the ``invisible'' social connections that bind people to places. Community cohesion during school redistricting processes is often an important community goal~\cite{bridges2016eden}, yet such cohesion can also exacerbate segregation.

Approximately 13 years ago, WS/FCS created custom planning units ("segments") to delineate neighborhoods and other important social infrastructure across the community, comprising just under 800 total units. These units were also defined, in part, in terms of current school attendance boundaries. This led to an important question: should these segments be used to redraw boundaries now? On the one hand, they encode valuable community information that may help produce rezoning plans that are more socially acceptable. On the other hand, using them could implicitly bias towards the status quo and perhaps limit the nature of improvements that redistricting plans may produce. An early conversation with the project's steering committee---an advisory body comprised of school board members and other community leaders---revealed a desire to ``not use old pieces to make something new'' (paraphrase). To leverage the useful information that segments contained while creating more degrees of freedom for possible redistricting plans, the project team decided to cut segments by Census Block Groups, to respect natural and human-made boundaries. This led to just over 3,000 pieces---many of which were small and oddly shaped. To achieve compact planning units (which would, in turn, help yield more compact boundaries in redistricting plans), the team developed a greedy algorithm to merge pieces in order to increase compactness of the individual planning units used for re-assignment. This process yielded approximately 1,500 units per school level.

To further increase the likelihood of regularly-shaped boundaries (in addition to the above merging process, and the compactness objective described in the prior section), the team also cut edges in the adjacency graph (representing planning unit adjacencies) used to impose contiguity constraints. In particular, for any planning unit A with at least three neighbors, the team would identify the neighbor with the shortest shared boundary with A and cut the corresponding edge in the adjacency graph (so long as doing so would not introduce island units, i.e., units with no connections in the graph). This reduced the likelihood of ``weakly-contiguous'' boundaries with units only connected through thin regions.

\subsubsection{Driving distances}
The paper uses a local instance of the OpenRouteService~\cite{ors2022} (compiled in Fall 2021) to estimate driving distances (in miles) from each student's address to each school in the district. This service is free to download and use, and yielded nearly identical results (on average) as driving distances computed using the Google Maps API on a small random sample of students.

\subsection{Stakeholder Engagement}
\label{subsect:stakeholder}
Engagement with two key groups of stakeholders---i) community members (parents, school staff, and students), and ii) district staff---was essential for planning and model development. Engagement occurred largely between April and December 2024 to understand hopes and concerns about the project and possible changes to attendance boundaries. Over 10,000 community members participated, with the vast majority (over 8,400) responding to a survey (others also participated in facilitated listening sessions). A more thorough investigation of these insights will be covered in future work. For the purposes of this study, the team turned to the survey, which included a question asking respondents which factors are most important to them when deciding which schools students should attend. Options included: 1) ``a school that is close to home'', 2) ``the social and economic diversity of students at the school'', 3) ``advancing to middle/high school with most of the same classmates'', 4) ``a school that isn’t too overcrowded or under-enrolled'', 5) ``the school’s academic performance grade'' (a coarse measure of school quality), and 6) ``the availability of specialty programs and extracurricular activities at the school''. While these were pre-set options that respondents ranked in the survey, the team found them to also be prevalent themes in the open-ended feedback community members provided through both the survey and community listening sessions. 

Perhaps unsurprisingly, respondents tended to rank 5) highest: consistently, it received the most first-place rankings when aggregating across respondents from different schools. This reflects prior research highlighting how school ratings impact where families live and send their kids to schools~\cite{zhang2008flight,black1999housing,kane2005housing}. In many schools, 6) was also highly ranked. However, both 5) and 6) were not factors the team felt could be fairly incorporated into the redistricting model. Building in school ratings or a school's programs either as part of the objective function or constraints could implicitly suggest that these are fixed attributes of schools---when in reality, they are features that can change through district decisions (like which programs are offered at the school; school improvement initiatives; and even who attends certain schools---which could activate Parent-Teacher organizations and other parent advocacy to attract new programs and resources to schools~\cite{murrayPTA}). School safety was another theme that emerged from open-ended feedback, but was not included as a ranking option because of its use in prior settings as a coded way of expressing racialized preferences for schools~\cite{billingham2020safety}---and was not factored in as a modeling priority for similar reasons as school performance and programs (as well as the additional challenge of lacking standardized data to describe school safety). Therefore, the team focused on designing the model's objectives and constraints around the other factors.

Engagement with district staff also played an essential role in informing model development. For example, upon seeing an initial set of boundaries and results, one district team member wondered if it might be possible to still achieve gains on integration and travel while additionally requiring that feeder pattern splits do not increase. This led the team to incorporate this new constraint into the model as an additional configuration parameter. (A feeder pattern split occurs when a school's population splits into multiple schools at the next level, like a middle school feeding into multiple high schools.) A constraint preventing schools from becoming more segregated than they currently are in status quo also emerged from iterative dialogue with the district (described in the next section). The team initially started with a school capacity objective in addition to the capacity constraint, but dropped that after discussions illuminating the complexity of determining an ideal school capacity.

\subsubsection{Weights}
\label{sec:weights}

Survey responses were used to determine the weights assigned to the demographic balance, distance, and feeder pattern objective function terms described in Section~\ref{subsect:objective_function}. For this purpose, the team considered the relative rankings of each respondent among ranking factors 1), 2), and 3), which were interpreted as encoding preferences for reducing driving distance, improving demographic balance, and reducing split feeder patterns, respectively.

The survey responses included some representational disparities. For instance, the district student population is roughly equal parts White, Black, and Hispanic, but more than half of survey responses reported White. To measure and correct for such disparities in the weight calculations, the team compared survey responses to racial demographic data from the district (a demographic breakdown by income may have been more immediately relevant for this project's focus on socioeconomic integration, but data on the income breakdown of district families was not available). A small fraction of respondents expressed ``Prefer not to say'' or ``Prefer to self-describe'' on the survey. For purposes of comparing to the district data, which does not have these categories, the race of these respondents was estimated by averaging the known racial demographics of their affiliated schools.

Weights are intended to reflect per-school preferences. School affiliations were gathered from other survey questions about student/child attendance or staff/other affiliation, depending on the stakeholder group. Some respondents had many school affiliations, for instance, staff members who support many schools. If a respondent had more than 3 affiliations, their contributions to the weights were normalized so their total influence was equal to a respondent with 3 affiliations.

The weights $w_{s,obj}$ were then calculated as the fraction of first place preferences for objective $obj$ (among the three options under consideration) among respondents affiliated with school $s$, reweighted by race to correct for representational disparities. For instance, if 50\% of a school is Black but only 30\% of survey respondents affiliated with that school are Black, then their preferences are upweighted to contribute 50\% of that school's weights. The weights are normalized by construction, so that $\sum_{obj} w_{s,obj} = 1$.

\subsection{Experimental Settings}
\label{subsect:experiment}

\begin{table}[t]
\footnotesize
\centering
\caption{This table includes set sizes and parameter values applied to the Winston-Salem case study. Parameter values were selected by consulting prior work~\cite{gillani2023redrawing} and through discussions with district partners.}
\begin{tabularx}{\columnwidth}{l X}
\toprule
Sets \& Parameters & Set, Set Size, or Parameter Value \\
\midrule
$\mathcal{L}$ & \{1, 2, 3\} Elementary, Middle, and High \\
$|\mathcal{SH}_1|, |\mathcal{SH}_2|, |\mathcal{SH}_3| $ & 41, 16, 10   \\
$|\mathcal{N}_1|, |\mathcal{N}_2|, |\mathcal{N}_3| $ & 22523, 11621, 16434    \\
$|\mathcal{P}_1|, |\mathcal{P}_2|, |\mathcal{P}_3| $ & 1506, 1495, 1507  See Section~\ref{subsect:data}.   \\
$\mathcal{O}$ & \{distance, balance, compact, feeder\}  These are chosen after  discussions with the district. \\
$\mu_n^\text{ratio}$ & 1 for all students  \\
$\epsilon$ & 1 \\
$\lambda$ & 0.15  \\
$o_s^{\max} \enspace o_s^{\min}$ & Each school in the dataset has two values: the status-quo zoned number of students and the serviceable capacity. In some cases, the zoned number exceeds the serviceable number, which also occurs in this case study. The value of $o_s^{\max}$ is determined by taking the maximum of these two values, while $o_s^{\min}$ is obtained by taking the minimum and multiplying it by 0.85.\\
\bottomrule
\end{tabularx}
\label{tab:parameters}
\vspace{-10pt}
\end{table}

As discussed, community and district input played an integral role in shaping the model. This section evaluates different versions of the model to offer an empirical investigation of how the different objectives and inputs impact how the district might redraw attendance boundaries. Table~\ref{tab:parameters} summarizes the parameters used for the different evaluated configurations. Overall, the team evaluated seven experiments, each redrawing boundaries simultaneously across all three school levels in $\mathcal{L}$. The Status-Quo (\textbf{SQ}) experiment simply evaluates the current zoning configuration and does not involve any boundary redrawing; this serves as a baseline for the model. The four single-objective runs, \textbf{S-TR}, \textbf{S-DB}, \textbf{S-C}, and \textbf{S-FB}, each targets a distinct objective: Travel Distance Ratio, Demographic Balancing between Lower-SES students and others, Compactness, and Feeder Pattern, respectively. These runs correspond to the four objectives \eqref{eq:objective_distance}, \eqref{eq:objective_balance}, \eqref{eq:objective_feeder}, and \eqref{eq:objective_compactness} introduced in Section~\ref{subsect:objective_function}.  The Boundary Capacity objective is excluded due to its functional similarity to Constraint~\eqref{formulation:constraint_school_capacity}, which also considers school capacity. Finally, there are two multi-objective runs, \textbf{N-NW} and \textbf{N-SW}, which seek to optimize all four aforementioned objectives simultaneously. \textbf{N-NW} uses uniform weights, while \textbf{N-SW} applies school-specific weights $w_{s, obj}$ to reflect local priorities. These configurations were informed by the district's preferences: in some cases, the district was interested in understanding what a map that only optimizes for one goal might look like (and how much it might still make progress on other goals); in other cases, they sought maps that simultaneously optimize for all of the goals. These single-objective runs are also used for calibration (see \Cref{sec:calibration}).

As introduced in Section~\ref{subsect:cp_constraints}, additional constraints can be incorporated as needed for a specific district when solving SABR. In this case study, when evaluating \textbf{S-DB}, \textbf{N-NW}, and \textbf{N-SW}, an additional  Constraint~\eqref{formulation:constraint_dissimilairty_bound} is added. As discussed earlier, it imposes a limit on $d_s$, preventing a school from deviating further from its original ratio of the specific group's population to the total population, except when the initial ratio already falls within some margin value $\lambda$ (which is set to 15\% for the experiments). The intuition for this constraint is an effort to prevent further segregating schools, but allowing fluctuations within some acceptable window, (similar, again, to the range of acceptable enrollments prescribed by lower courts after desegregation litigation like~\cite{swann1971charlotte}). Here, $\bar{o}_s$ and $\bar{o}^\text{g}_s$ represent the number of students and the students in group $g$ zoned to school $s$ according to the status-quo zoning.

\begin{equation}
\label{formulation:constraint_dissimilairty_bound}
\resizebox{0.75\columnwidth}{!}{
$
    d_s 
    \leq 
    \max\left(
        \lambda,
        \cfrac{\bar{o}^\text{g}_s}{\bar{o}_s}
        -
        \cfrac{|\mathcal{G}^l|}{|\mathcal{N}^l|}
    \right)
    \quad \forall l \in \mathcal{L}, s \in \mathcal{SH}^l
$
}
\end{equation}

To evaluate the results, this study does not report the raw objective values directly, as they are more difficult to interpret. Instead, it uses the following metrics:
\begin{itemize}[label=---]
    \item Average Student Driving Distance: Reflects the daily experience of students and parents. This metric relates to the Travel Distance Ratio objective.
    \item District Dissimilarity: While the model optimizes  $\mathbf{f}_{balance}$, this metric modifies $\mathbf{f}_{balance-d}$ by removing the weights to provide a district-wide measure of demographic balance, see Equation~\eqref{eq:dissimilarity_district_evaluate}. The value of this equation always falls between 0 and 1, where 0 indicates no segregation between the two groups. This is the popular dissimilarity measure of segregation, introduced in~\cite{massey1988segregation}, and applied here to measure segregation between lower- and non-lower-SES students.
    \item \# Feeder Splits: Sames as the feeder pattern objective.
    \item \# Rezoned Lower-SES students, any students, and planning units: Measures the scale of rezoning, offering a clear view of how extensively student assignments and geographic boundaries change.
\end{itemize}

\begin{equation}
\label{eq:dissimilarity_district_evaluate}
\resizebox{0.85\columnwidth}{!}{
$
    \text{District Dissimilarity} = \cfrac{1}{2}
    \sum\limits_{s \in \mathcal{SH}^l}
    \bigg|
    \cfrac{o^\text{g}_s}{|\mathcal{G}^l|}
    -
    \cfrac{o_s - o^\text{g}_s}{|\mathcal{N}^l| - |\mathcal{G}^l|}
    \bigg|
$
}
\end{equation}

This study uses Python 3.11 for all programming tasks. The optimization model runs on CP-SAT, a solver within Google OR-Tools, an open-source library \cite{cpsat}. Note that picking CP-SAT for modeling SABR is important, as the model is intended for direct use by school districts, and therefore should not depend on commercial solvers. Except for \textbf{SQ}, all runs apply the CP model with a time limit of 10 hours. The status-quo zoning is treated as a warm-start for each configuration. The solver terminates early if it finds an optimal solution; otherwise, it continues until the time limit is reached. Additionally, each experiment runs 30 times with 30 different random seeds for the CP-SAT solver to test whether variations in the seed lead to significantly different results.

\begin{table*}[t]
\caption{Results from solving the optimization model with a 10-hour time limit. Each experiment ran 30 times using different random seeds. If any run reached the optimal solution, the table reports results related to the optimal solution. Otherwise, it reports the average result across the 30 runs after 10 hours, with standard errors in parentheses.}
\begin{subtable}{0.9\textwidth}
\small
\caption{Computational Results for Level 1: Elementary Schools. Standard errors appear in parentheses.}
\resizebox{\textwidth}{!}{
\begin{tabular}{
    l r 
    r r r 
    r r r
}
\toprule
&&
\multicolumn{3}{c}{\textbf{Measurements Related to Objectives}} &
\multicolumn{3}{c}{\textbf{\# [\%] Rezoned}} 
\\ \cmidrule(lr){3-5} \cmidrule(lr){6-8}
\makecell[l]{Exp.} &
\makecell[c]{Optimal} &
\makecell[r]{Average Student \\ Driving Distance \\   (miles)} & 
\makecell[r]{District \\ Dissimilarity} & 
\makecell[r]{\# Feeder \\ Splits} &
\makecell[r]{Lower-SES \\ Students} & 
\makecell[r]{Any \\  Students} & 
\makecell[r]{Planning \\ Units} \\
\midrule

\textbf{SQ} & - & 2.20 \phantom{(.0)}(-) & 0.62 \phantom{(.0)}(-) & 59 \phantom{(.0)}(-) & - & - & -\\

\textbf{S-TR} & \ding{55} & 2.05 (0.00) & 0.59 (0.00) & 90 (0.63) & 2246 (28.74) [19.81\%] & 4285 (31.16) [19.03\%] & 344 (1.91) [22.81\%]\\

\textbf{S-DB} & \ding{55} & 2.16 (0.00) & 0.54 (0.00) & 105 (0.46) & 2043 (16.34) [18.02\%] & 4491 (23.46) [19.94\%] & 385 (1.66) [25.59\%]\\

\textbf{S-C} & \ding{55} & 2.16 (0.00) & 0.61 (0.00) & 83 (0.30) & 1988 (19.20) [17.54\%] & 3299 (28.81) [14.65\%] & 291 (1.64) [19.31\%]\\

\textbf{S-FP} & \ding{51} & 2.20 \phantom{(.0)}(-) & 0.59 \phantom{(.0)}(-) & 47 \phantom{(.0)}(-) & 1690 \phantom{(0.0)}(-) [14.91\%] & 3195 \phantom{(0.0)}(-) [14.19\%] & 295 \phantom{(.0)}(-) [19.59\%]\\

\textbf{M-NW} & \ding{55} & 2.12 (0.00) & 0.57 (0.00) & 63 (0.99) & 1911 (25.64) [16.86\%] & 3641 (47.16) [16.16\%] & 314 (2.52) [20.83\%]\\

\textbf{M-SW} & \ding{55} & 2.12 (0.00) & 0.57 (0.00) & 65 (1.68) & 1936 (30.17) [17.08\%] & 3678 (51.85) [16.33\%] & 316 (3.04) [20.98\%]\\

\bottomrule
\end{tabular}
}
\end{subtable}
\par  \vspace{0.05cm}
\begin{subtable}{0.9\textwidth}
\caption{Computational Results for Level 2: Middle Schools. Standard errors appear in parentheses.}
\resizebox{\textwidth}{!}{
\begin{tabular}{
    l r 
    r r r 
    r r r
}
\toprule
&&
\multicolumn{3}{c}{\textbf{Measurements Related to Objectives}} &
\multicolumn{3}{c}{\textbf{\# [\%] Rezoned}} 
\\ \cmidrule(lr){3-5} \cmidrule(lr){6-8}
\makecell[l]{Exp.} &
\makecell[c]{Optimal} &
\makecell[r]{Average Student \\ Driving Distance \\   (miles)} & 
\makecell[r]{District \\ Dissimilarity} & 
\makecell[r]{\# Feeder \\ Splits} &
\makecell[r]{Lower-SES \\ Students} & 
\makecell[r]{Any \\  Students} & 
\makecell[r]{Planning \\ Units} \\
\midrule

\textbf{SQ} & - & 3.35 \phantom{(.0)}(-) & 0.57 \phantom{(.0)}(-) & 38 \phantom{(.0)}(-) & - & - & -\\

\textbf{S-TR} & \ding{55} & 3.04 (0.00) & 0.56 (0.00) & 45 (0.25) & 1689 (10.49) [28.40\%] & 2671 (22.89) [22.98\%] & 353 (2.33) [23.64\%]\\

\textbf{S-DB} & \ding{55} & 3.40 (0.01) & 0.41 (0.00) & 54 (0.29) & 1982 (27.45) [33.33\%] & 3693 (30.04) [31.78\%] & 486 (3.86) [32.54\%]\\

\textbf{S-C} & \ding{55} & 3.12 (0.00) & 0.57 (0.00) & 41 (0.28) & 1488 (15.53) [25.02\%] & 2502 (18.18) [21.53\%] & 359 (1.88) [24.03\%]\\

\textbf{S-FP} & \ding{51} & 3.21 \phantom{(.0)}(-) & 0.55 \phantom{(.0)}(-) & 23 \phantom{(.0)}(-) & 1238 \phantom{(0.0)}(-) [20.82\%] & 1996 \phantom{(0.0)}(-) [17.18\%] & 286 \phantom{(.0)}(-) [19.13\%]\\

\textbf{M-NW} & \ding{55} & 3.07 (0.00) & 0.53 (0.00) & 32 (0.40) & 1579 (14.31) [26.55\%] & 2584 (19.28) [22.23\%] & 339 (2.15) [22.67\%]\\

\textbf{M-SW} & \ding{55} & 3.08 (0.00) & 0.53 (0.00) & 32 (0.46) & 1566 (22.82) [26.33\%] & 2570 (30.95) [22.11\%] & 339 (3.39) [22.68\%]\\
\bottomrule
\end{tabular}
}
\end{subtable}
\par \vspace{0.05cm}
\begin{subtable}{0.9\textwidth}
\caption{Computational Results for Level 3: High Schools. Standard errors appear in parentheses.}
\resizebox{\textwidth}{!}{
\begin{tabular}{
    l r 
    r r r 
    r r r
}
\toprule
&&
\multicolumn{3}{c}{\textbf{Measurements Related to Objectives}} &
\multicolumn{3}{c}{\textbf{\# [\%] Rezoned}} 
\\ \cmidrule(lr){3-5} \cmidrule(lr){6-8}
\makecell[l]{Exp.} &
\makecell[c]{Optimal} &
\makecell[r]{Average Student \\ Driving Distance \\   (miles)} & 
\makecell[r]{District \\ Dissimilarity} & 
\makecell[r]{\# Feeder \\ Splits} &
\makecell[r]{Lower-SES \\ Students} & 
\makecell[r]{Any \\  Students} & 
\makecell[r]{Planning \\ Units} \\
\midrule

\textbf{SQ} & - & 4.03 \phantom{(.0)}(-) & 0.41 \phantom{(.0)}(-) & - & - & - & -\\

\textbf{S-TR} & \ding{55} & 3.75 (0.00) & 0.53 (0.00) & - & 1942 (10.06) [24.75\%] & 3882 (22.95) [23.62\%] & 385 (2.32) [25.58\%]\\

\textbf{S-DB} & \ding{55} & 4.12 (0.00) & 0.36 (0.00) & - & 2232 (36.04) [28.45\%] & 3774 (64.22) [22.97\%] & 370 (5.64) [24.57\%]\\

\textbf{S-C} & \ding{55} & 3.91 (0.01) & 0.50 (0.01) & - & 1568 (68.30) [19.99\%] & 2796 (104.21) [17.01\%] & 300 (8.25) [19.91\%]\\

\textbf{S-FP} & \ding{51} & 3.89 \phantom{(.0)}(-) & 0.47 \phantom{(.0)}(-) & - & 1479 \phantom{(0.0)}(-) [18.86\%] & 2853 \phantom{(0.0)}(-) [17.36\%] & 311 \phantom{(.0)}(-) [20.64\%]\\

\textbf{M-NW} & \ding{55} & 3.94 (0.00) & 0.41 (0.00) & - & 1143 (29.86) [14.57\%] & 2123 (58.75) [12.92\%] & 236 (5.27) [15.64\%]\\

\textbf{M-SW} & \ding{55} & 3.94 (0.00) & 0.41 (0.00) & - & 1216 (41.16) [15.50\%] & 2218 (67.16) [13.49\%] & 243 (6.23) [16.10\%]\\

\bottomrule
\end{tabular}
}
\end{subtable}
\label{tab:results}
\end{table*}

\begin{figure*}[!ht]
    \centering
    \begin{subfigure}[h]{0.26\textwidth}
        \includegraphics[width=\textwidth]{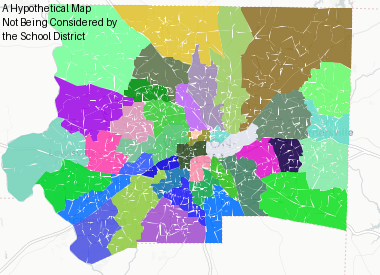}
        \caption{SQ: Elementary Schools}
    \end{subfigure}
    \begin{subfigure}[h]{0.26\textwidth}   \includegraphics[width=\textwidth]{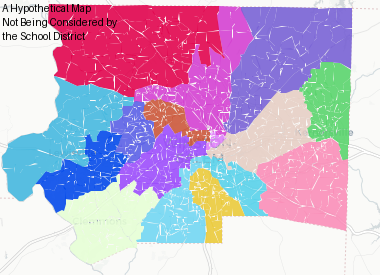}
        \caption{SQ: Middle Schools}
    \end{subfigure}
    \begin{subfigure}[h]{0.26\textwidth}\includegraphics[width=\textwidth]{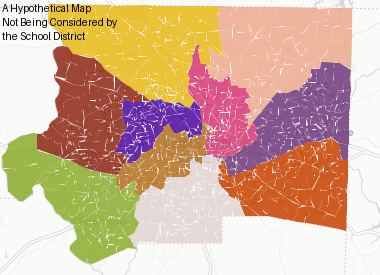}
        \caption{SQ: High Schools}
    \end{subfigure}
    
    \vspace{0.05cm}
    
    \begin{subfigure}[h]{0.26\textwidth}
        \includegraphics[width=\textwidth]{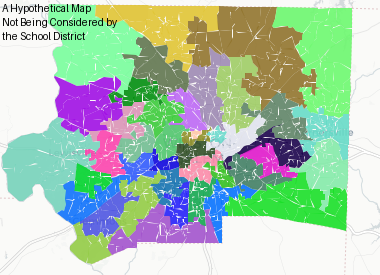}
        \caption{S-DB: Elementary Schools}
    \end{subfigure}
    \begin{subfigure}[h]{0.26\textwidth}    \includegraphics[width=\textwidth]{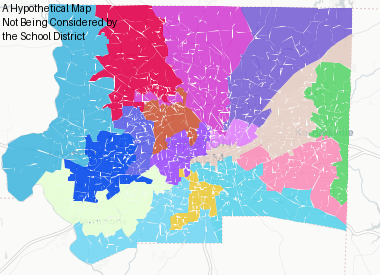}
        \caption{S-DB: Middle Schools}
    \end{subfigure}
    \begin{subfigure}[h]{0.26\textwidth}
        \includegraphics[width=\textwidth]{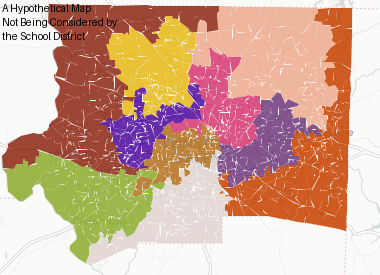}
        \caption{S-DB: High Schools}
    \end{subfigure}

    \vspace{0.05cm}
    \begin{subfigure}[h]{0.26\textwidth}
        \includegraphics[width=\textwidth]{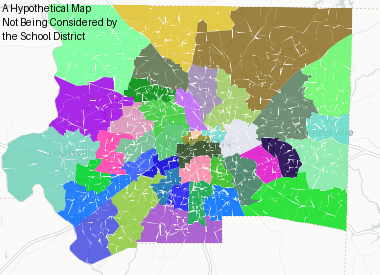}
        \caption{S-C: Elementary Schools}
        \label{subfig:S-C-Elementary}
    \end{subfigure}
    \begin{subfigure}[h]{0.26\textwidth}
        \includegraphics[width=\textwidth]{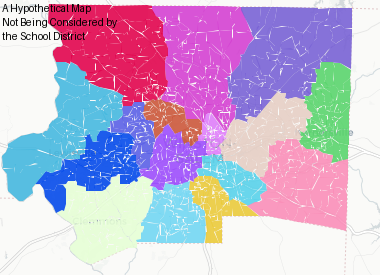}
        \caption{S-C: Middle Schools}
        \label{subfig:S-C-Middle}
    \end{subfigure}
    \begin{subfigure}[h]{0.26\textwidth}
        \includegraphics[width=\textwidth]{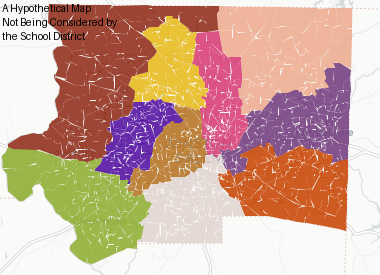}
        \caption{S-C: High Schools}
        \label{subfig:S-C-High}
    \end{subfigure}
    
    \vspace{0.05cm}
        
    \begin{subfigure}[h]{0.26\textwidth}
        \includegraphics[width=\textwidth]{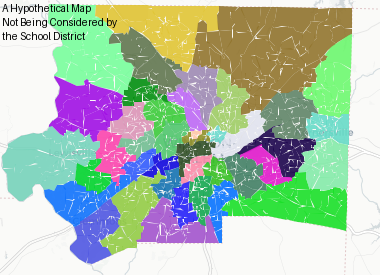}
        \caption{M-SW: Elementary Schools}
    \end{subfigure}
    \begin{subfigure}[h]{0.26\textwidth}
        \includegraphics[width=\textwidth]{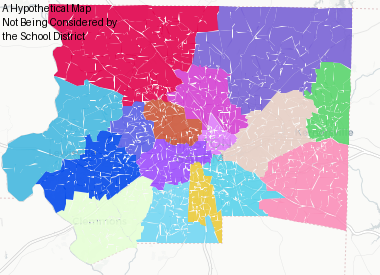}
        \caption{M-SW: Middle Schools}
    \end{subfigure}
    \begin{subfigure}[h]{0.26\textwidth}
        \includegraphics[width=\textwidth]{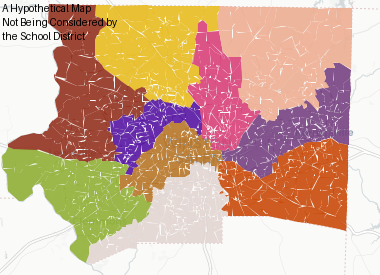}
        \caption{M-SW: High Schools}
    \end{subfigure}
\caption{Zoning Visualizations for Experiments SQ, S-DB, S-C, and M-SW.}
\Description{}
\label{fig:maps_part_1}
\end{figure*}
\subsection{Computational Results}
Table~\ref{tab:results} presents the computational results from the seven experiments. For each experiment, the ``Optimal'' column indicates whether any of the 30 runs successfully found an optimal solution. If so, the results from the optimal run are reported. Otherwise, the table shows the mean of each metric across the 30 runs, with standard error in parentheses. As shown in the table, all metrics have small standard errors, indicating that the CP model performs consistently and is stable for this particular task. Figures~\ref{fig:maps_part_1} provide visualizations of some zonings (to reserve space), allowing for direct comparison across different experiments. If an experiment does not find the optimal solution, the mean $Objective(\mathcal{L})$ is calculated across all runs, and its corresponding map is shown.

Table~\ref{tab:results} reveals a clear and expected pattern across all single-objective experiments: results from \textbf{S-TR}, \textbf{S-DB}, and \textbf{S-FP} all achieve the best performance on the metric corresponding to the objective that particular configuration optimized for. In particular, while the average student driving distance in \textbf{S-TR} shows only a slight improvement over \textbf{SQ}, the large number of students involved makes this a potentially-significant gain for the district. The \textbf{S-DB} experiment seeks to reduce SES segregation between schools. To do so, the model reassigns a larger number of students. As shown in the table, \textbf{S-DB} produces the most extensive boundary changes at the middle school level, followed by elementary and high, respectively. Consequently, it reassigns the highest number of lower-SES students at both the middle and high school levels. Interestingly, the baseline district-wide dissimilarity at the high school level under \textbf{SQ} is already lower than the elementary and middle levels, likely due to the small number of high schools (10) and their subsequently larger attendance boundaries. Even so, \textbf{S-DB High} achieves a modest relative reduction in dissimilarity of approximately 12\% (0.41 to 0.36). The tradeoff, however, is an increase over \textbf{SQ} in average driving distance.

Visual inspection of Figures~\ref{subfig:S-C-Elementary}, \ref{subfig:S-C-Middle}, and \ref{subfig:S-C-High}, each representing zoning plans optimized for compactness (\textbf{S-C}), confirms the effectiveness of this objective. Qualitatively speaking, compared to other maps, these maps exhibit less irregular shapes. Results from \textbf{S-FP} are particularly interesting, as it is the only experiment that reaches optimality. In fact, all 30 runs find the optimal solution in approximately 10 minutes. This efficiency likely results from the relatively small number of possible feeder pattern combinations (especially when combined with travel and capacity constraints, which limit the space of possible re-assignments). Moreover, at the elementary school level, the status-quo zoning, which is the warm-start of the model, only includes 59 feeder patterns, while \textbf{S-FP} reduces this number to just 47 in the optimal solution. Both \textbf{S-C} and \textbf{S-FP} are closely tied to geographical factors, and Table~\ref{tab:results} shows they generally require less drastic boundary changes compared to \textbf{S-TR} and \textbf{S-DB}. This may be because the status-quo zoning, designed by policymakers in response to (likely) community desires for stable feeder patterns, already incorporates careful consideration of geographic features.

The final two experiments are multi-objective runs. Most of the time, they improve status-quo zoning metrics without requiring drastic boundary changes. However, as expected, their gains on each metric are typically lower than the corresponding gains on those metrics under the single objective runs. This reflects trade-offs that the multi-objective models tries to balance.

\begin{figure*}[!ht]
    \centering
    
    \begin{subfigure}[h]{0.28\textwidth}
        \includegraphics[width=\textwidth]{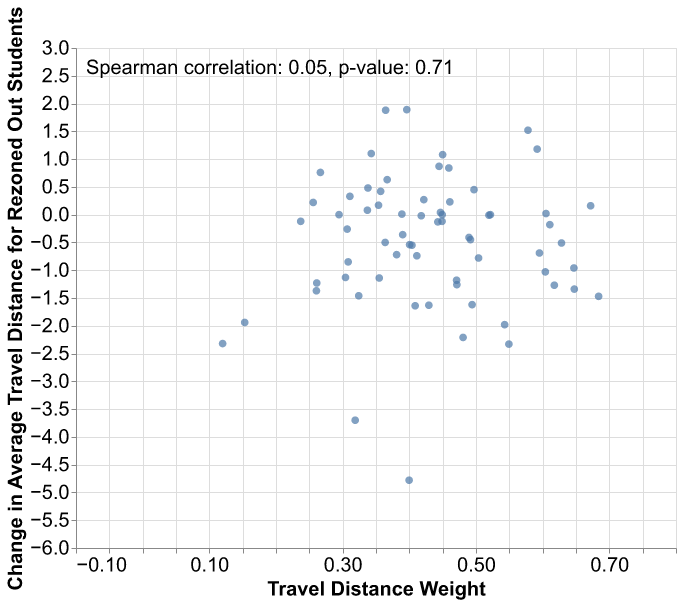}
        \caption{M-NW: Travel Distance}
    \end{subfigure}
    \begin{subfigure}[h]{0.28\textwidth}
        \includegraphics[width=\textwidth]{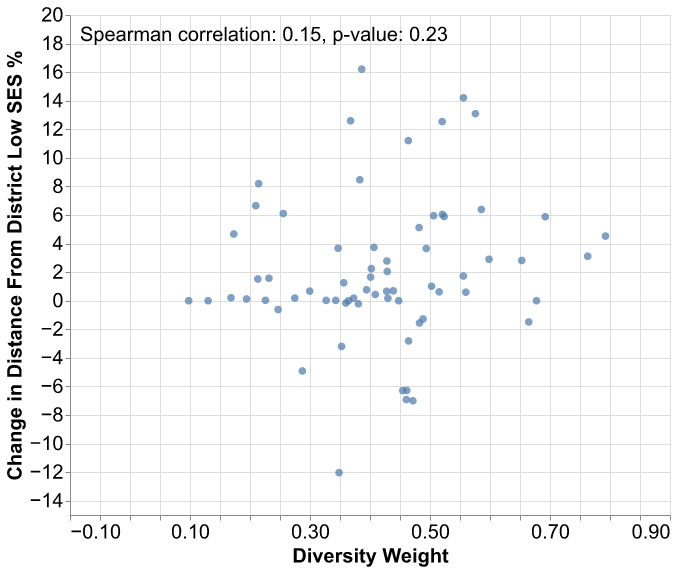}
        \caption{M-NW: Diversity}
    \end{subfigure}
    \begin{subfigure}[h]{0.28\textwidth}
        \includegraphics[width=\textwidth]{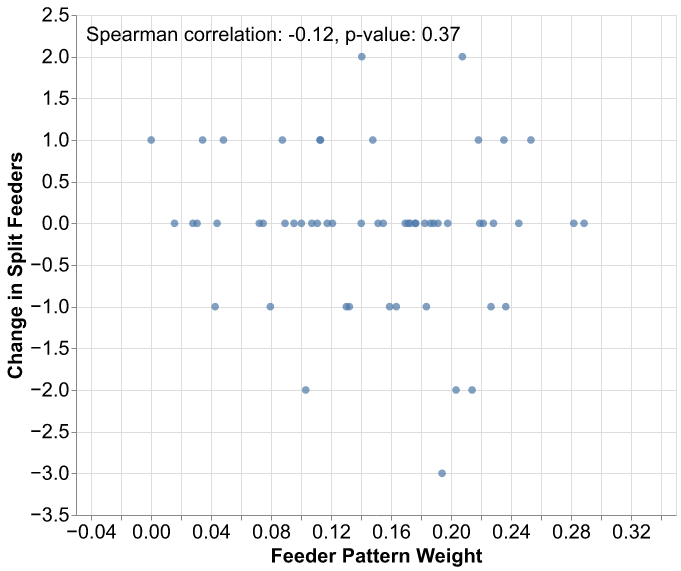}
        \caption{M-NW: Feeder Patterns}
    \end{subfigure}
    
    \vspace{0.35cm}

    \begin{subfigure}[h]{0.28\textwidth}
        \includegraphics[width=\textwidth]{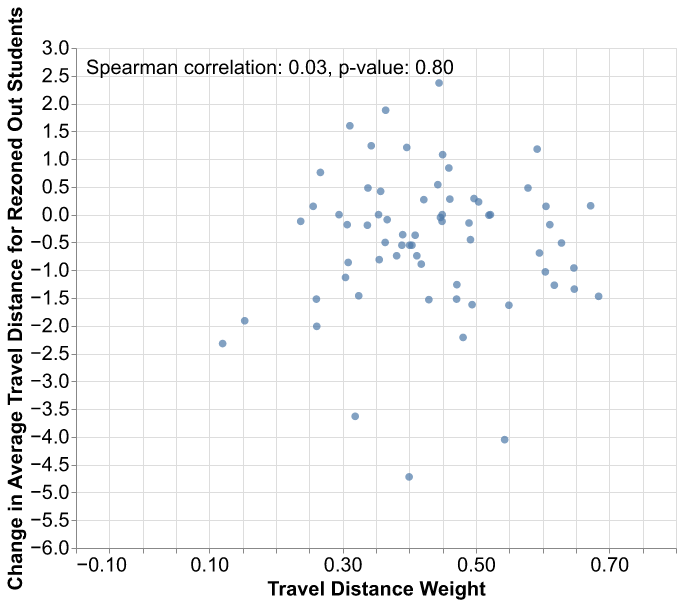}
        \caption{M-SW: Travel Distance}
    \end{subfigure}
    \begin{subfigure}[h]{0.28\textwidth}
        \includegraphics[width=\textwidth]{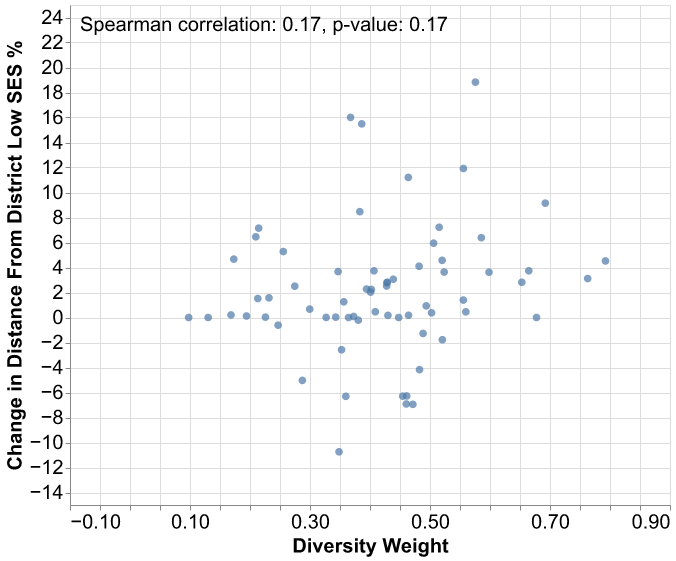}
        \caption{M-SW: Diversity}
    \end{subfigure}
    \begin{subfigure}[h]{0.28\textwidth}
        \includegraphics[width=\textwidth]{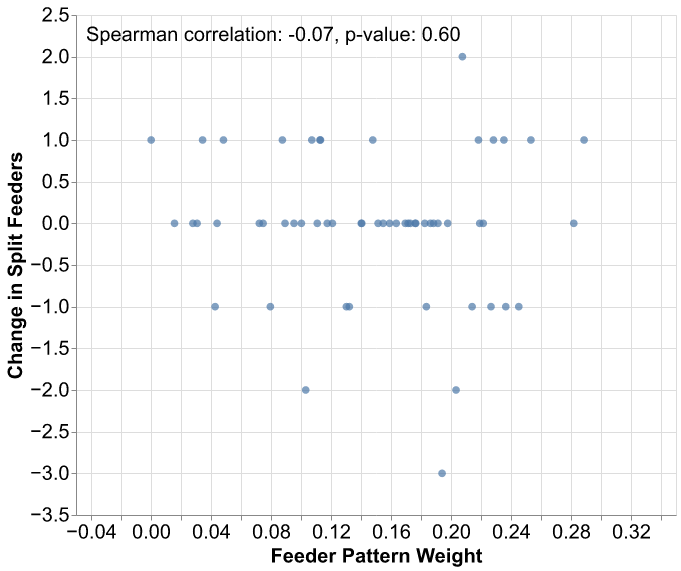}
        \caption{M-SW: Feeder Patterns}
    \end{subfigure}
\caption{Spearman Correlations Between Weights and Outcome Metrics ("Diversity" refers to SES integration).}
\Description{}
\label{fig:weight_charts}
\end{figure*}
To investigate how incorporating survey-derived weights may have impacted results, Figure~\ref{fig:weight_charts} shows the Spearman correlation between the weights and result metrics from experiments \textbf{M-NW} and \textbf{M-SW} for each school in the district. The SES integration weights per school are compared to how much that school increases in SES integration post-rezoning (i.e., how much its percentage of zoned lower-SES students moves toward the district-level percentage for that school level): more positive changes imply more integration. The travel distance weights are compared with changes in average travel distance for students rezoned out of a given school. The feeder pattern weights are compared with changes in the number of split feeder patterns for a given school.

The correlations between survey-derived weights and their corresponding outcomes under \textbf{M-NW} are small and statistically insignificant. This is to be expected given \textbf{M-NW} did not incorporate the survey-derived weights, and instead, assigned equal weights to each objective. However, correlations between weights and outcomes under \textbf{M-SW}---the model that incorporated survey input---are also small and statistically insignificant, suggesting that weights did not consistently shape outcomes across schools under this model. There are many possible reasons for this: a notable one is that the sociogeography of the district and other constraints in the model may have inherently limited the extent to which boundary changes might reflect the preferences of respondents. For example, respondents from schools with high concentrations of lower-SES students often tended to express the highest preferences for integration; however, many of these schools are clustered together in the district's urban core, which limits the extent to which boundary changes might actually foster integration at these schools---especially when the travel objective/constraint are imposed. This highlights an important limitation of redistricting that the project team is currently exploring how to address, through additional dialogue with these communities. Strategies beyond boundary changes may include re-imagining school choice policies and addressing barriers to exercising choice, for example, through the strategic siting of new magnet programs~\cite{hawkins2018sanantonio} and controlled choice initiatives~\cite{alves1987controlled}.

Crucially, the ``null'' result of incorporating survey-derived weights should not be viewed as an indication that community input has no bearing on redistricting outcomes. For both philosophical and practical reasons (e.g., at the level of individual schools in different parts of the district), community voices are important to incorporate into the redistricting algorithm's decision-making process, as they may have different impacts in different geographic contexts and under different constraint values. A thorough evaluation of the different conditions under which weights might impact redistricting outcomes is beyond the scope of this paper, but serves as an exciting and important direction for future work.

\subsection{Converging on Final Maps for Deployment}
\label{subsect:deployment}

The results in the preceding section offer an empirical evaluation of different redistricting scenarios. The project team discussed several such scenarios and others with district leadership, including ones with varying travel time constraints. The ability to quickly incorporate new constraints (like the aforementioned Feeder pattern constraint suggested by district collaborators) and vary values for existing constraints to regenerate maps enabled rapid policy exploration and evaluation. In the end, the project's steering committee decided to release three draft maps---A, B, and C---to the community for feedback. Map A was closest to \textbf{S-DB}; Map B was closest to \textbf{M-SW}, with a more flexible travel constraint (which enabled slightly larger gains in integration, and slightly smaller average decreass in driving times); and Map C was closest to \textbf{M-SW} as presented in the preceding section. Importantly, both B and C incorporated the additional feeder constraint mentioned earlier. The committee decided to release these maps because they each offered a relevant reference point for the project: Map A illustrated what might be achieved by optimizing only for the project's namesake (more SES-integrated schools); Map B showed how optimizing for all goals with a slightly more flexible travel constraint might bring improvements across all metrics compared to the status quo; and Map C served as a more conservative version of Map B, which would involve fewer changes (and therefore, more muted gains, particularly on SES integration). 

Noting the perennial challenge and controversy surrounding attendance boundary changes---especially in a district that had not meaningfully changed them in a generation---it was particularly important to the district to present options that would demonstrate progress across all relevant metrics (hence, the decision to include Maps B and C, and add constraints---as discussed earlier---that seek to guarantee such progress). The team also worked with the district to ensure no racial/ethnic or SES groups were burdened or harmed more than others under the selected plans. This particularly involved ensuring that increases in driving distances were not unfairly experienced by one group over another, like in prior eras of desegregation, when Black families largely shouldered the burden of ``busing''~\cite{delmont2016busing}. Fortunately, the team did not observe such systematic disparate impacts (and future models may more explicitly impose fairness constraints to help guarantee fair outcomes in different demographic and geographic contexts). As noted earlier, an enduring unfairness is that many schools with the highest concentrations of lower-SES students also expressed the strongest preferences for integration---but they would experience minimal impacts under alternative boundaries. Working closely with these communities to identify alternative pathways to more integrated schools will be a primary area of focus in the coming months.

SABR fortunately surfaced rezoning proposals capable of advancing several community-centered goals, in an otherwise complex space of possible rezonings. In the coming months, the project team looks forward to analyzing and iterating upon community responses to these maps, and sharing findings in future studies.

\section{Discussion}
\label{sect:discussion}

This study presents and evaluates a novel algorithmic redistricting framework, shaped by both policymaker and community input, for redrawing school attendance boundaries. For researchers and practitioners in education, computer science, and optimization, the study illustrates both methodological and practical considerations required to carry out such work in a large school district with a mix of ideological perspectives, and offers a \href{https://github.com/Plural-Connections/EAAMO-2025-Redrawing-School-Attendance-Boundaries}{\textcolor{blue}{toolbox}} that can easily be adapted to other settings.

There are a number of limitations in this work, many of which may serve as fertile ground for future research and practice. One is the model's inability to solve to optimality across most instances, due to the large-scale nature of the redistricting. Future work that builds upon~\cite{ozel2025community} to constrain the space of possible rezonings via more intelligent data preparation may improve solution quality. Another limitation is the extent to which the team explored how survey-derived weights might affect results. While the relatively straight-forward incorporation of survey inputs as weights showed little impact on outcomes in the WS/FCS context, computing weights differently, and/or including them alongside different constraint values, may yield impacts. Finally, there remain opportunities to more comprehensively evaluate different model parameters and their impacts. Prior work offers some insight~\cite{gillani2023redrawing}.

An important question that often accompanies algorithm-supported policymaking is: should algorithms be deciding humans' futures? Throughout this study, the team has attempted to acknowledge and counter the false dichotomy implicit in this question---i.e., that either algorithms or humans are responsible for unilaterally deciding the future---by 1) listening to community members and policymakers and bringing their voices into the design and inputs of the model, and 2) ensuring that algorithmically-proposed policies do not place undue burden or harm on any particular demographic group. Furthermore, the team has positioned algorithms as tools for identifying first (not final) drafts of boundary proposals. In this way, the team has sought to embody a spirit of intellectual humility~\cite{porter2017humility} as a guiding principal: a belief that knowledge should be gleaned from varied sources to produce proposals that advance the project's goals as best as possible, and then, those proposals should be subjected to evaluation by the people whose lives they are likely to impact. Still, given the high-stakes nature of this policymaking process, school districts will continue to benefit from new ideas about how algorithmic and human competencies may complement one another to draft policies that are just, fair, and advance quality of life for everyone. The team hopes this study is a useful building block for others engaging in this challenging, yet essential, work.

\begin{acks}
The authors thank the WT Grant Foundation, Overdeck Family Foundation, the US Department of Education's ``Fostering Diverse Schools'' grant, and the National Science Foundation (Grant \#2112533) for funding that made this work possible. They also extend their appreciation to various staff and community members in Winston-Salem/Forsyth County Schools for their ongoing support and collaboration.
\end{acks}
\bibliographystyle{ACM-Reference-Format}
\bibliography{references}


\begin{thebibliography}{47}


\ifx \showCODEN    \undefined \def \showCODEN     #1{\unskip}     \fi
\ifx \showISBNx    \undefined \def \showISBNx     #1{\unskip}     \fi
\ifx \showISBNxiii \undefined \def \showISBNxiii  #1{\unskip}     \fi
\ifx \showISSN     \undefined \def \showISSN      #1{\unskip}     \fi
\ifx \showLCCN     \undefined \def \showLCCN      #1{\unskip}     \fi
\ifx \shownote     \undefined \def \shownote      #1{#1}          \fi
\ifx \showarticletitle \undefined \def \showarticletitle #1{#1}   \fi
\ifx \showURL      \undefined \def \showURL       {\relax}        \fi
\providecommand\bibfield[2]{#2}
\providecommand\bibinfo[2]{#2}
\providecommand\natexlab[1]{#1}
\providecommand\showeprint[2][]{arXiv:#2}

\bibitem[swa(nd)]%
        {swann1971charlotte}
 \bibinfo{year}{n.d.}\natexlab{}.
\newblock \bibinfo{title}{Swann v. Charlotte-Mecklenburg Board of Education}.
\newblock \bibinfo{howpublished}{Oyez}.
\newblock
\urldef\tempurl%
\url{https://www.oyez.org/cases/1970/281}
\showURL{%
\tempurl}


\bibitem[Allman et~al\mbox{.}(2022)]%
        {allman2022}
\bibfield{author}{\bibinfo{person}{Maxwell Allman}, \bibinfo{person}{Itai Ashlagi}, \bibinfo{person}{Irene Lo}, \bibinfo{person}{Juliette Love}, \bibinfo{person}{Katherine Mentzer}, \bibinfo{person}{Lulabel Ruiz-Setz}, {and} \bibinfo{person}{Henry O'Connell}.} \bibinfo{year}{2022}\natexlab{}.
\newblock \showarticletitle{Designing School Choice for Diversity in the San Francisco Unified School District}. In \bibinfo{booktitle}{\emph{Proceedings of the 23rd ACM Conference on Economics and Computation}}. \bibinfo{publisher}{ACM}, \bibinfo{address}{Boulder, CO, USA}, \bibinfo{pages}{290--291}.
\newblock


\bibitem[Alves and Willie(1987)]%
        {alves1987controlled}
\bibfield{author}{\bibinfo{person}{Michael~J. Alves} {and} \bibinfo{person}{Charles~V. Willie}.} \bibinfo{year}{1987}\natexlab{}.
\newblock \showarticletitle{Controlled choice assignments: A new and more effective approach to school desegregation}.
\newblock \bibinfo{journal}{\emph{Urban Review}} \bibinfo{volume}{19}, \bibinfo{number}{2} (\bibinfo{year}{1987}), \bibinfo{pages}{67--88}.
\newblock
\href{https://doi.org/10.1007/BF01121341}{doi:\nolinkurl{10.1007/BF01121341}}


\bibitem[Billingham et~al\mbox{.}(2020)]%
        {billingham2020safety}
\bibfield{author}{\bibinfo{person}{Chase Billingham}, \bibinfo{person}{Shelley~M. Kimelberg}, \bibinfo{person}{Sarah Faude}, {and} \bibinfo{person}{Matthew~O. Hunt}.} \bibinfo{year}{2020}\natexlab{}.
\newblock \showarticletitle{In Search of a Safe School: Racialized Perceptions of Security and the School Choice Process}.
\newblock \bibinfo{journal}{\emph{The Sociological Quarterly}} \bibinfo{volume}{61}, \bibinfo{number}{3} (\bibinfo{year}{2020}), \bibinfo{pages}{474--499}.
\newblock


\bibitem[Black(1999)]%
        {black1999housing}
\bibfield{author}{\bibinfo{person}{Sandra~E. Black}.} \bibinfo{year}{1999}\natexlab{}.
\newblock \showarticletitle{{Do Better Schools Matter? Parental Valuation of Elementary Education}}.
\newblock \bibinfo{journal}{\emph{The Quarterly Journal of Economics}} \bibinfo{volume}{114}, \bibinfo{number}{2} (\bibinfo{year}{1999}), \bibinfo{pages}{577--599}.
\newblock


\bibitem[Bouzarth et~al\mbox{.}(2018)]%
        {bouzarth2018assigning}
\bibfield{author}{\bibinfo{person}{Elizabeth~L Bouzarth}, \bibinfo{person}{Richard Forrester}, \bibinfo{person}{Kevin~R Hutson}, {and} \bibinfo{person}{Lattie Reddoch}.} \bibinfo{year}{2018}\natexlab{}.
\newblock \showarticletitle{Assigning students to schools to minimize both transportation costs and socioeconomic variation between schools}.
\newblock \bibinfo{journal}{\emph{Socio-Economic Planning Sciences}}  \bibinfo{volume}{64} (\bibinfo{year}{2018}), \bibinfo{pages}{1--8}.
\newblock


\bibitem[Bridges(2016)]%
        {bridges2016eden}
\bibfield{author}{\bibinfo{person}{Kim Bridges}.} \bibinfo{year}{2016}\natexlab{}.
\newblock \bibinfo{title}{Eden Prairie Public Schools: Adapting to Demographic Change in the Suburbs}.
\newblock \bibinfo{howpublished}{The Century Foundation}.
\newblock
\urldef\tempurl%
\url{https://tcf.org/content/report/eden-prairie-public-schools/}
\showURL{%
\tempurl}


\bibitem[Castro et~al\mbox{.}(2024)]%
        {castro2024drawn}
\bibfield{author}{\bibinfo{person}{Andrene Castro}, \bibinfo{person}{Genevieve Siegel-Hawley}, \bibinfo{person}{Kimberly Bridges}, {and} \bibinfo{person}{Shenita~E Williams}.} \bibinfo{year}{2024}\natexlab{}.
\newblock \showarticletitle{Drawn into policy: A systematic review of school rezoning rationales, processes, and outcomes}.
\newblock \bibinfo{journal}{\emph{Review of Educational Research}} \bibinfo{volume}{94}, \bibinfo{number}{4} (\bibinfo{year}{2024}), \bibinfo{pages}{539--583}.
\newblock


\bibitem[Chetty et~al\mbox{.}(2020)]%
        {chetty2018race}
\bibfield{author}{\bibinfo{person}{Raj Chetty}, \bibinfo{person}{Nathaniel Hendren}, \bibinfo{person}{Maggie~R Jones}, {and} \bibinfo{person}{Sonya~R Porter}.} \bibinfo{year}{2020}\natexlab{}.
\newblock \showarticletitle{Race and economic opportunity in the United States: An intergenerational perspective}.
\newblock \bibinfo{journal}{\emph{The Quarterly Journal of Economics}} \bibinfo{volume}{135}, \bibinfo{number}{2} (\bibinfo{year}{2020}), \bibinfo{pages}{711--783}.
\newblock


\bibitem[Chetty et~al\mbox{.}(2022)]%
        {chetty2022socialcapitalII}
\bibfield{author}{\bibinfo{person}{Raj Chetty}, \bibinfo{person}{Matthew~O. Jackson}, \bibinfo{person}{Theresa Kuchler}, {and} \bibinfo{person}{Johannes et~al. Stroebel}.} \bibinfo{year}{2022}\natexlab{}.
\newblock \showarticletitle{Social capital II: determinants of economic connectedness}.
\newblock \bibinfo{journal}{\emph{Nature}}  \bibinfo{volume}{608} (\bibinfo{year}{2022}), \bibinfo{pages}{122--134}.
\newblock


\bibitem[Clarke and Surkis(1968)]%
        {clarke1968}
\bibfield{author}{\bibinfo{person}{S Clarke} {and} \bibinfo{person}{J Surkis}.} \bibinfo{year}{1968}\natexlab{}.
\newblock \showarticletitle{An operations research approach to racial desegregation of school systems}.
\newblock \bibinfo{journal}{\emph{Socio-Economic Planning Sciences}} \bibinfo{volume}{1}, \bibinfo{number}{3} (\bibinfo{year}{1968}), \bibinfo{pages}{259--272}.
\newblock


\bibitem[Dantec and Fox(2015)]%
        {le-dantec2015strangers}
\bibfield{author}{\bibinfo{person}{Christopher A.~Le Dantec} {and} \bibinfo{person}{Sarah Fox}.} \bibinfo{year}{2015}\natexlab{}.
\newblock \showarticletitle{Strangers at the Gate: Gaining Access, Building Rapport, and Co-Constructing Community-Based Research}. In \bibinfo{booktitle}{\emph{Proceedings of the 18th ACM Conference on Computer Supported Cooperative Work \& Social Computing}} (Vancouver, BC, Canada) \emph{(\bibinfo{series}{CSCW '15})}. \bibinfo{address}{New York, NY, USA}, \bibinfo{pages}{1348--1358}.
\newblock
\href{https://doi.org/10.1145/2675133.2675147}{doi:\nolinkurl{10.1145/2675133.2675147}}


\bibitem[Delmont(2016)]%
        {delmont2016busing}
\bibfield{author}{\bibinfo{person}{Matthew~F. Delmont}.} \bibinfo{year}{2016}\natexlab{}.
\newblock \bibinfo{booktitle}{\emph{{Why Busing Failed: Race, Media, and the National Resistance to School Desegregation}}}. Vol.~\bibinfo{volume}{42}.
\newblock \bibinfo{publisher}{University of California Press}.
\newblock


\bibitem[Diamond and Wright(1987)]%
        {diamond1987multiobjective}
\bibfield{author}{\bibinfo{person}{James~T Diamond} {and} \bibinfo{person}{Jeff~R Wright}.} \bibinfo{year}{1987}\natexlab{}.
\newblock \showarticletitle{Multiobjective analysis of public school consolidation}.
\newblock \bibinfo{journal}{\emph{Journal of Urban Planning and Development}} \bibinfo{volume}{113}, \bibinfo{number}{1} (\bibinfo{year}{1987}), \bibinfo{pages}{1--18}.
\newblock


\bibitem[Dube and Clark(2016)]%
        {dube2016beyond}
\bibfield{author}{\bibinfo{person}{Matthew~P Dube} {and} \bibinfo{person}{Jesse~T Clark}.} \bibinfo{year}{2016}\natexlab{}.
\newblock \showarticletitle{Beyond the circle: Measuring district compactness using graph theory}. In \bibinfo{booktitle}{\emph{Northeast Political Science Association Conference}}. \bibinfo{publisher}{NPSA}, \bibinfo{address}{Boston, Massachuset, USA}.
\newblock


\bibitem[Fiel(2013)]%
        {fiel2013boundaries}
\bibfield{author}{\bibinfo{person}{Jeremy~E. Fiel}.} \bibinfo{year}{2013}\natexlab{}.
\newblock \showarticletitle{{Decomposing School Resegregation: Social Closure, Racial Imbalance, and Racial Isolation}}.
\newblock \bibinfo{journal}{\emph{American Sociological Review}} \bibinfo{volume}{78}, \bibinfo{number}{5} (\bibinfo{year}{2013}).
\newblock


\bibitem[Garms and Starke(2023)]%
        {wsfoundation}
\bibfield{author}{\bibinfo{person}{Layla Garms} {and} \bibinfo{person}{Shamika Starke}.} \bibinfo{year}{2023}\natexlab{}.
\newblock \bibinfo{title}{Rooted in Race: Our Community’s History of School Integration}.
\newblock \bibinfo{howpublished}{The Winston-Salem Foundation}.
\newblock
\urldef\tempurl%
\url{https://www.wsfoundation.org/blog/rooted-in-race}
\showURL{%
\tempurl}


\bibitem[Gillani et~al\mbox{.}(2023a)]%
        {gillani2023redrawing}
\bibfield{author}{\bibinfo{person}{Nabeel Gillani}, \bibinfo{person}{Doug Beeferman}, \bibinfo{person}{Christine Vega-Pourheydarian}, \bibinfo{person}{Cassandra Overney}, \bibinfo{person}{Pascal Van~Hentenryck}, {and} \bibinfo{person}{Deb Roy}.} \bibinfo{year}{2023}\natexlab{a}.
\newblock \showarticletitle{{Redrawing attendance boundaries to promote racial and ethnic diversity in elementary schools}}.
\newblock \bibinfo{journal}{\emph{Educational Researcher}} \bibinfo{volume}{52}, \bibinfo{number}{6} (\bibinfo{year}{2023}), \bibinfo{pages}{348--364}.
\newblock


\bibitem[Gillani et~al\mbox{.}(2023b)]%
        {gillani2023air}
\bibfield{author}{\bibinfo{person}{Nabeel Gillani}, \bibinfo{person}{Cassandra Overney}, \bibinfo{person}{Claire Schuch}, {and} \bibinfo{person}{Kumar Chandra}.} \bibinfo{year}{2023}\natexlab{b}.
\newblock \bibinfo{title}{Fostering More Integrated Schools Through Community-Driven, Machine-Informed Rezoning}.
\newblock \bibinfo{howpublished}{Essay for the American Institutes of Research, Integration and Equity 2.0: New Approaches for a New Era}.
\newblock
\urldef\tempurl%
\url{https://www.air.org/integration-and-equity-2-0-essays/community-approaches-and-perspectives#fostering}
\showURL{%
\tempurl}


\bibitem[GIScience(2022)]%
        {ors2022}
\bibfield{author}{\bibinfo{person}{GIScience}.} \bibinfo{year}{2022}\natexlab{}.
\newblock \bibinfo{title}{Open Route Service}.
\newblock
\urldef\tempurl%
\url{https://github.com/GIScience/openrouteservice}
\showURL{%
\tempurl}


\bibitem[Guan et~al\mbox{.}(2025)]%
        {guan2025rwc}
\bibfield{author}{\bibinfo{person}{Hongzhao Guan}, \bibinfo{person}{Nabeel Gillani}, \bibinfo{person}{Tyler Simko}, \bibinfo{person}{Jasmine Mangat}, {and} \bibinfo{person}{Pascal Van~Hentenryck}.} \bibinfo{year}{2025}\natexlab{}.
\newblock \showarticletitle{Contextual Stochastic Optimization for School Desegregation Policymaking}. In \bibinfo{booktitle}{\emph{Proceedings of the AAAI Conference on Artificial Intelligence}}, Vol.~\bibinfo{volume}{39}. \bibinfo{publisher}{AAAI Press}, \bibinfo{address}{Washington, DC, USA}, \bibinfo{pages}{28024--28032}.
\newblock


\bibitem[Gurnee and Shmoys(2021)]%
        {gurnee2021fairmandering}
\bibfield{author}{\bibinfo{person}{Wes Gurnee} {and} \bibinfo{person}{David~B. Shmoys}.} \bibinfo{year}{2021}\natexlab{}.
\newblock \showarticletitle{{Fairmandering: A column generation heuristic for fairness-optimized political districting}}.
\newblock \bibinfo{journal}{\emph{arXiv: 2103.11469}} (\bibinfo{year}{2021}), \bibinfo{pages}{88--99}.
\newblock


\bibitem[Harwell and LeBeau(2010)]%
        {harwell2010frl}
\bibfield{author}{\bibinfo{person}{Michael Harwell} {and} \bibinfo{person}{Brandon LeBeau}.} \bibinfo{year}{2010}\natexlab{}.
\newblock \showarticletitle{{Student Eligibility for a Free Lunch as an SES Measure in Education Research}}.
\newblock \bibinfo{journal}{\emph{Educational Researcher}} \bibinfo{volume}{39}, \bibinfo{number}{2} (\bibinfo{year}{2010}).
\newblock


\bibitem[Hawkins(2018)]%
        {hawkins2018sanantonio}
\bibfield{author}{\bibinfo{person}{Beth Hawkins}.} \bibinfo{year}{2018}\natexlab{}.
\newblock \bibinfo{title}{78207: America’s Most Radical School Integration Experiment}.
\newblock \bibinfo{howpublished}{The 74 Million}.
\newblock
\urldef\tempurl%
\url{https://www.the74million.org/article/78207-americas-most-radical-school-integration-experiment/}
\showURL{%
\tempurl}


\bibitem[Holloway et~al\mbox{.}(1975)]%
        {holloway1975interactive}
\bibfield{author}{\bibinfo{person}{Charles~A Holloway}, \bibinfo{person}{Donald~A Wehrung}, \bibinfo{person}{Michael~P Zeitlin}, {and} \bibinfo{person}{Rosser~T Nelson}.} \bibinfo{year}{1975}\natexlab{}.
\newblock \showarticletitle{An interactive procedure for the school boundary problem with declining enrollment}.
\newblock \bibinfo{journal}{\emph{Operations Research}} \bibinfo{volume}{23}, \bibinfo{number}{2} (\bibinfo{year}{1975}), \bibinfo{pages}{191--206}.
\newblock


\bibitem[Jacobson(2023)]%
        {jacobson2023fostering}
\bibfield{author}{\bibinfo{person}{Linda Jacobson}.} \bibinfo{year}{2023}\natexlab{}.
\newblock \bibinfo{title}{Feds Award Millions to School Districts to Address ‘Tricky’ Issue of Integration}.
\newblock \bibinfo{howpublished}{The 74 Million}.
\newblock
\urldef\tempurl%
\url{https://www.the74million.org/article/feds-award-millions-to-school-districts-to-address-tricky-issue-of-integration/}
\showURL{%
\tempurl}


\bibitem[Johnson(2011)]%
        {johnson2011desegregation}
\bibfield{author}{\bibinfo{person}{Rucker~C. Johnson}.} \bibinfo{year}{2011}\natexlab{}.
\newblock \showarticletitle{{Long-run Impacts of School Desegregation \& School Quality on Adult Attainments}}.
\newblock \bibinfo{journal}{\emph{NBER Working Paper No. 16664}} (\bibinfo{year}{2011}).
\newblock


\bibitem[Kane et~al\mbox{.}(2006)]%
        {kane2005housing}
\bibfield{author}{\bibinfo{person}{Thomas~J Kane}, \bibinfo{person}{Stephanie~K Riegg}, {and} \bibinfo{person}{Douglas~O Staiger}.} \bibinfo{year}{2006}\natexlab{}.
\newblock \showarticletitle{School quality, neighborhoods, and housing prices}.
\newblock \bibinfo{journal}{\emph{American law and economics review}} \bibinfo{volume}{8}, \bibinfo{number}{2} (\bibinfo{year}{2006}), \bibinfo{pages}{183--212}.
\newblock


\bibitem[Kennedy(2007)]%
        {pics_kennedy}
\bibfield{author}{\bibinfo{person}{J. Kennedy}.} \bibinfo{year}{2007}\natexlab{}.
\newblock \bibinfo{title}{Parents Involved in Community Schools v. Seattle School District}.
\newblock \bibinfo{howpublished}{https://www.law.cornell.edu/supct/pdf/05-908P.ZC1}.
\newblock
\newblock
\shownote{551 US 701}.


\bibitem[Lemberg and Church(2000)]%
        {lemberg2000school}
\bibfield{author}{\bibinfo{person}{David~S Lemberg} {and} \bibinfo{person}{Richard~L Church}.} \bibinfo{year}{2000}\natexlab{}.
\newblock \showarticletitle{The school boundary stability problem over time}.
\newblock \bibinfo{journal}{\emph{Socio-Economic Planning Sciences}} \bibinfo{volume}{34}, \bibinfo{number}{3} (\bibinfo{year}{2000}), \bibinfo{pages}{159--176}.
\newblock


\bibitem[Massey and Denton(1988)]%
        {massey1988segregation}
\bibfield{author}{\bibinfo{person}{D.~S. Massey} {and} \bibinfo{person}{N.~A. Denton}.} \bibinfo{year}{1988}\natexlab{}.
\newblock \showarticletitle{{The Dimensions of Residential Segregation}}.
\newblock \bibinfo{journal}{\emph{{Social Forces}}} \bibinfo{volume}{67}, \bibinfo{number}{2} (\bibinfo{year}{1988}), \bibinfo{pages}{281--315}.
\newblock


\bibitem[Mehrotra et~al\mbox{.}(1998)]%
        {mehrotra1998contiguity}
\bibfield{author}{\bibinfo{person}{Anuj Mehrotra}, \bibinfo{person}{Ellis~L. Johnson}, {and} \bibinfo{person}{George~L. Nemhauser}.} \bibinfo{year}{1998}\natexlab{}.
\newblock \showarticletitle{{An Optimization Based Heuristic for Political Districting}}.
\newblock \bibinfo{journal}{\emph{Management Science}} \bibinfo{volume}{44}, \bibinfo{number}{8} (\bibinfo{year}{1998}), \bibinfo{pages}{1021--1166}.
\newblock


\bibitem[Murray et~al\mbox{.}(2019)]%
        {murrayPTA}
\bibfield{author}{\bibinfo{person}{B. Murray}, \bibinfo{person}{T. Domina}, \bibinfo{person}{L. Renzulli}, {and} \bibinfo{person}{R. Boylan}.} \bibinfo{year}{2019}\natexlab{}.
\newblock \showarticletitle{Civil Society Goes to School: Parent-Teacher Associations and the Equality of Educational Opportunity}.
\newblock \bibinfo{journal}{\emph{RSF: The Russell Sage Foundation Journal of the Social Sciences}} \bibinfo{volume}{5}, \bibinfo{number}{3} (\bibinfo{year}{2019}), \bibinfo{pages}{41--63}.
\newblock


\bibitem[OR-Tools(2025)]%
        {cpsat}
\bibfield{author}{\bibinfo{person}{Google OR-Tools}.} \bibinfo{year}{2025}\natexlab{}.
\newblock \bibinfo{title}{CP-SAT Solver}.
\newblock
\urldef\tempurl%
\url{https://developers.google.com/optimization/cp/cp_solver}
\showURL{%
\tempurl}


\bibitem[Overney et~al\mbox{.}(2025)]%
        {overney2025boundarease}
\bibfield{author}{\bibinfo{person}{Cassandra Overney}, \bibinfo{person}{Cassandra Moe}, \bibinfo{person}{Alvin Chang}, {and} \bibinfo{person}{Nabeel Gillani}.} \bibinfo{year}{2025}\natexlab{}.
\newblock \showarticletitle{BoundarEase: Fostering Constructive Community Engagement to Inform More Equitable Student Assignment Policies}.
\newblock \bibinfo{journal}{\emph{Proceedings of the ACM on Human-Computer Interaction}} \bibinfo{volume}{9}, \bibinfo{number}{CSCW2} (\bibinfo{date}{April} \bibinfo{year}{2025}), \bibinfo{pages}{Article CSCW040, 37 pages}.
\newblock
\href{https://doi.org/10.1145/3710938}{doi:\nolinkurl{10.1145/3710938}}


\bibitem[Ozel et~al\mbox{.}(2025)]%
        {ozel2025community}
\bibfield{author}{\bibinfo{person}{Aysu Ozel}, \bibinfo{person}{Karen Smilowitz}, {and} \bibinfo{person}{Lila~KS Goldstein}.} \bibinfo{year}{2025}\natexlab{}.
\newblock \showarticletitle{Community-Engaged School District Design: A Stream-Based Approach}.
\newblock \bibinfo{journal}{\emph{Operations Research}}  \bibinfo{volume}{Articles in Advance} (\bibinfo{year}{2025}).
\newblock


\bibitem[Pathak and Shi(2021)]%
        {pathak2017demand}
\bibfield{author}{\bibinfo{person}{Parag~A Pathak} {and} \bibinfo{person}{Peng Shi}.} \bibinfo{year}{2021}\natexlab{}.
\newblock \showarticletitle{How well do structural demand models work? Counterfactual predictions in school choice}.
\newblock \bibinfo{journal}{\emph{Journal of Econometrics}} \bibinfo{volume}{222}, \bibinfo{number}{1} (\bibinfo{year}{2021}), \bibinfo{pages}{161--195}.
\newblock


\bibitem[Porter and Schumann(2018)]%
        {porter2017humility}
\bibfield{author}{\bibinfo{person}{Tenelle Porter} {and} \bibinfo{person}{Karina Schumann}.} \bibinfo{year}{2018}\natexlab{}.
\newblock \showarticletitle{Intellectual humility and openness to the opposing view}.
\newblock \bibinfo{journal}{\emph{Self and Identity}} \bibinfo{volume}{17}, \bibinfo{number}{2} (\bibinfo{year}{2018}), \bibinfo{pages}{139--162}.
\newblock


\bibitem[Quick(2016)]%
        {quick2016cps}
\bibfield{author}{\bibinfo{person}{Kimberly Quick}.} \bibinfo{year}{2016}\natexlab{}.
\newblock \bibinfo{title}{Chicago Public Schools: Ensuring Diversity in Selective Enrollment and Magnet Schools}.
\newblock \bibinfo{howpublished}{The Century Foundation}.
\newblock
\urldef\tempurl%
\url{https://tcf.org/content/report/chicago-public-schools/}
\showURL{%
\tempurl}


\bibitem[Reardon et~al\mbox{.}(2019a)]%
        {reardon2019geography}
\bibfield{author}{\bibinfo{person}{Sean~F Reardon}, \bibinfo{person}{Demetra Kalogrides}, {and} \bibinfo{person}{Kenneth Shores}.} \bibinfo{year}{2019}\natexlab{a}.
\newblock \showarticletitle{The Geography of Racial/Ethnic Test Score Gaps}.
\newblock \bibinfo{journal}{\emph{Amer. J. Sociology}} \bibinfo{volume}{124}, \bibinfo{number}{4} (\bibinfo{year}{2019}), \bibinfo{pages}{1164--1221}.
\newblock


\bibitem[Reardon et~al\mbox{.}(2019b)]%
        {reardon2018testgaps}
\bibfield{author}{\bibinfo{person}{Sean~F Reardon}, \bibinfo{person}{Demetra Kalogrides}, {and} \bibinfo{person}{Kenneth Shores}.} \bibinfo{year}{2019}\natexlab{b}.
\newblock \showarticletitle{The geography of racial/ethnic test score gaps}.
\newblock \bibinfo{journal}{\emph{Amer. J. Sociology}} \bibinfo{volume}{124}, \bibinfo{number}{4} (\bibinfo{year}{2019}), \bibinfo{pages}{1164--1221}.
\newblock


\bibitem[Schools(2025)]%
        {cms2025review}
\bibfield{author}{\bibinfo{person}{Charlotte-Mecklenburg Schools}.} \bibinfo{year}{2025}\natexlab{}.
\newblock \bibinfo{title}{Comprehensive Data Review}.
\newblock \bibinfo{howpublished}{Charlotte-Mecklenburg Schools}.
\newblock
\urldef\tempurl%
\url{https://www.cmsk12.org/ComprehensiveReview}
\showURL{%
\tempurl}


\bibitem[Simko(2024)]%
        {simko2024school}
\bibfield{author}{\bibinfo{person}{Tyler Simko}.} \bibinfo{year}{2024}\natexlab{}.
\newblock \showarticletitle{School desegregation by redrawing district boundaries}.
\newblock \bibinfo{journal}{\emph{Scientific Reports}} \bibinfo{volume}{14}, \bibinfo{number}{1} (\bibinfo{year}{2024}), \bibinfo{pages}{22097}.
\newblock


\bibitem[Smilowitz and Keppler(2020)]%
        {smilowitz2020}
\bibfield{author}{\bibinfo{person}{Karen Smilowitz} {and} \bibinfo{person}{Samantha Keppler}.} \bibinfo{year}{2020}\natexlab{}.
\newblock \showarticletitle{On the use of operations research and management in public education systems}.
\newblock \bibinfo{journal}{\emph{Pushing the boundaries: Frontiers in impactful OR/OM research}} (\bibinfo{year}{2020}), \bibinfo{pages}{84--105}.
\newblock


\bibitem[Sutcliffe et~al\mbox{.}(1984)]%
        {sutcliffe1984goal}
\bibfield{author}{\bibinfo{person}{Charles Sutcliffe}, \bibinfo{person}{John Board}, {and} \bibinfo{person}{Paul Cheshire}.} \bibinfo{year}{1984}\natexlab{}.
\newblock \showarticletitle{Goal programming and allocating children to secondary schools in Reading}.
\newblock \bibinfo{journal}{\emph{Journal of the Operational Research Society}} \bibinfo{volume}{35}, \bibinfo{number}{8} (\bibinfo{year}{1984}), \bibinfo{pages}{719--730}.
\newblock


\bibitem[Van~Hentenryck et~al\mbox{.}(1992)]%
        {van1992constraint}
\bibfield{author}{\bibinfo{person}{Pascal Van~Hentenryck}, \bibinfo{person}{Helmut Simonis}, {and} \bibinfo{person}{Mehmet Dincbas}.} \bibinfo{year}{1992}\natexlab{}.
\newblock \showarticletitle{Constraint satisfaction using constraint logic programming}.
\newblock \bibinfo{journal}{\emph{Artificial intelligence}} \bibinfo{volume}{58}, \bibinfo{number}{1-3} (\bibinfo{year}{1992}), \bibinfo{pages}{113--159}.
\newblock


\bibitem[Zhang(2008)]%
        {zhang2008flight}
\bibfield{author}{\bibinfo{person}{Haifeng Zhang}.} \bibinfo{year}{2008}\natexlab{}.
\newblock \showarticletitle{{White Flight in the Context of Education: Evidence from South Carolina}}.
\newblock \bibinfo{journal}{\emph{Journal of Geography}} \bibinfo{volume}{107}, \bibinfo{number}{6} (\bibinfo{year}{2008}), \bibinfo{pages}{236--245}.
\newblock


\end{thebibliography}
\end{document}